# Detecting intraday financial market states using temporal clustering

D. HENDRICKS*, T. GEBBIE and D. WILCOX

School of Computer Science and Applied Mathematics, University of the Witwatersrand
Johannesburg, WITS 2050, South Africa

We propose the application of a high-speed maximum likelihood clustering algorithm to detect temporal financial market states, using correlation matrices estimated from intraday market microstructure features. We first determine the *ex-ante* intraday temporal cluster configurations to identify market states, and then study the identified temporal state features to extract *state signature vectors* which enable online state detection. The *state signature vectors* serve as low-dimensional state descriptors which can be used in learning algorithms for optimal planning in the high-frequency trading domain. We present a feasible scheme for real-time intraday state detection from streaming market data feeds. This study identifies an interesting hierarchy of system behaviour which motivates the need for time-scale-specific state space reduction for participating agents.

## 1. Introduction

The financial market represents a prime example of an observable complex adaptive system. Many heterogeneous adaptive agents, such as traders, portfolio managers, market makers and regulatory authorities, interact non-linearly over time with each other and the electronic exchange, allowing for the emergence of complex behaviours beyond that expected based on intrinsic agent characteristics. Many authors have viewed financial markets through this lens, considering analogues with physical systems to formulate models which aid our understanding of observed system characteristics (see Arthur (1995), Arthur et al. (1997), Brock (1993), Hommes (2001), Wilcox and Gebbie (2014) and the references therein). Recent technological advances, accelerated by a highly competitive industry, have allowed for the efficient generation, storage and retrieval of financial data at micro time scales, providing a rich record of the price formation process as a laboratory for intensive study. The field of market microstructure developed to study the characteristics and behaviours of financial system dynamics at this scale (see O'Hara (1998), Madhavan (2000), Biais et al. (2005), Hinton (2007), Gençay et al. (2010), Baldovin et al. (2015) for a comprehensive discussion). In particular, as intraday trading and investment processes become increasingly automated, understanding the system dynamics at varying intraday time scales is critical for an efficient trajectory through the system to be mapped by participating agents.

This paper aims to use a physical analogy to the ferromagnetic Potts model at thermal equilibrium to describe object interactions, before deriving an unsupervised clustering algorithm, where both the number of clusters and configuration emerges from the data (Blatt et al. (1996, 1997), Wiseman et al. (1998), Giada and Marsili (2001)). Treating intraday time periods as objects, the algorithm will be used to identify intraday market states from observed market microstructure features. Although Marsili (2002) used a similar approach to classify days as states, the authors are unaware of another study which applies this technique to *intraday* period clustering using *multiple* features. In addition, a high-speed Parallel Genetic Algorithm (PGA) will be used for efficient computation of the cluster configurations, with absolute computation speeds conducive to overnight or even intraday recalibration of identified states (Hendricks et al. (2015)).

*Corresponding author. Email: dieter.hendricks@students.wits.ac.za

The results reveal an interesting hierarchy of system behaviour at different time scales. Statistically significant power-law fits to configuration characteristics suggest scale-invariant behaviour which may translate to persistent features in market states. In addition, the power-law fits yield different scaling exponents at the different time scales, suggesting that the system may be at criticality at each scale, possibly with different universality classes characterising behaviour (Dacorogna et al. (1996), Gabaix et al. (2003), Emmert-Streib and Dehmer (2010), Mastromatteo and Marsili (2011)). This motivates the importance of time-scale specific information when planning in this domain. Here we are considering a particular case of *calendar time* when investigating scale-related phenomena, however we note that there are alternative scales used to measure the financial system. There is a rich history in the literature which has aimed to directly model the *event time* foundations of market microstructure processes. The seminal work of Garman (1976), which used point processes to model order book events, forms the basis of many subsequent *event time* approaches to modelling transaction and quote data. An important extension of this view is the vector autoregressive model for trades and quotes developed by Hasbrouck (1988, 1991) and Engle and Russell (1998). A complementary approach introduces the concept of *intrinsic time*, which aims to measure trading opportunities in reference to specific features of traded stocks, for example, using the rate of trading to modify calendar or chronological time. These are discussed by Müller et al. (1995) and Derman (2002). The more recent use of Hawkes processes to model mutually-exciting order book events is an important return to the idea of viewing events as a foundational concept when modelling transactions and order book dynamics (Large (2000), Toke and Pomponio (2012), Bacry et al. (2015), Abergel and Jedidi (2015)).

Easley et al. (2012) introduce the *volume time* paradigm for high-frequency trading, with the clock ticking according to the number of events (proxied by trade volume) flowing through the system. This is a pragmatic attempt to reconcile the foundational event-based paradigm introduced by Garman (1976) with the wide use of chronological or calendar time. They argue that machines operate on a clock which is not chronological, but rather related to the number of cycles per instruction initiated by an event (Easley et al. (2012), Patterson and Hennessy (2013)). This allows one to measure time in terms of frequency of changes in information, as measured by trading volumes. When one considers the *complex event processing* paradigm which underpins many automated trading systems in financial markets (Adi et al. (2006)), one can appreciate the suitability of the event-based clock and the view that the calendar time clock is a legacy convenience from the low-frequency, human-trader-driven world. As the shift from human-driven to machine-driven trading dominates financial markets, the study of event-time-scale phenomena has become increasingly important and warrants further exploration.

While the identified market states reveal many interesting insights, trading agents would benefit from being able to detect online (or in real-time) which state they are *currently* in. We develop a novel technique which extracts the characteristic signature of market activity from each of the identified states, and uses this as the basis for an online state detection algorithm. In one application, this is used to construct 1-step transition probability matrices, which can be refined online and used in optimal planning algorithms.

This paper proceeds as follows: Section 2 describes a non-parametric clustering approach using a physical Potts model analogy. Section 3 uses the Potts model analogy to derive a maximum likelihood estimator for the optimal cluster configuration. Section 4 describes the idea of clustering time periods as objects in order to identify market states. Section 5 describes the parallel genetic algorithm for high-speed detection of the temporal cluster configuration using the maximum likelihood estimator. Section 6 describes an approach for online state detection. Section 7 discusses scale-invariant properties of the cluster configurations and how these may be exploited for efficient online state detection. Section 8 discusses the data used, workflow and results. Section 9 provides some concluding remarks and suggestions for further research.

## 2. Super-paramagnetic clustering

Blatt et al. (1996, 1997) and Wiseman et al. (1998) proposed a novel non-parametric clustering approach, based on an analogy to the ferromagnetic Potts model at thermal equilibrium. By assigning a Potts spin variable to each object and introducing a short-range distance-dependent ferromagnetic interaction field, regions of aligned spins emerge, which are analogous to groups of objects in the same cluster, where *spin alignment* suggests *object homogeneity* (Wang and Swendsen (1990)).

More formally, consider a $q$-state Potts model with spins $s_i = 1, ..., q$ for $i = 1, ..., N$, where $N$ is the total number of objects in the system. The cost function is given by the following Hamiltonian:

$$H = -\sum_{s_i, s_j \in S} J_{ij} \delta(s_i, s_j) \tag{1}$$

where the spins $s_i$ can take on $q$-states and the coupling of the $i^{th}$ and $j^{th}$ object are governed by $J_{ij}$. In the case of object clustering for a data sample, a candidate configuration is given by the set $\mathcal{S} = \{s_i\}_{i=1}^{n}$, where $s_i$ represents the cluster group index to which the $i^{th}$ object belongs. One can consider the coupling parameters $J_{ij}$ as being a function of the correlation coefficient $C_{ij}$ (Kullmann et al. (2000), Giada and Marsili (2001)). This is used to specify a distance function that is decreasing with distance between objects. If all the spins are related in this way, then each pair of spins is connected by some non-vanishing coupling $J_{ij} = J_{ij}(C_{ij})$. This allows one to interpret $s_i$ as a Potts spin in the Potts model Hamiltonian with $J_{ij}$ decreasing with the distance between objects (Blatt et al. (1996), Kullmann et al. (2000)). The case where there is only one cluster can be thought of as a ground state. As the system becomes more excited, it could break up into additional clusters. Each cluster would have specific Potts magnetisations, even though the nett magnetisation can be zero for the complete system. Generically, the correlation would then be both a function of time and temperature in order to encode both the evolution of clusters, as well as the hierarchy of clusters as a function of temperature. In the basic approach, one is looking for the lowest energy state that fits the data.

## 3. A maximum likelihood approach

In order to parameterise the model efficiently, one can choose to make an ansatz for the data generative function (Noh (2000)) and use this to develop a maximum-likelihood approach (Giada and Marsili (2001)), rather than explicitly solving the Potts Hamiltonian numerically (Blatt et al. (1996), Kullmann et al. (2000)). A number of authors have considered this approach for object clustering (McLachlan et al. (1996), Giada and Marsili (2001), Mungan and Ramasco (2010)), however we follow the proposition by Giada and Marsili (2001). A summary exposition will be presented here (as shown in Hendricks et al. (2015)), with a full derivation available in the Appendices.

According to the Noh (2000) ansatz, the generative model of the time series associated with the $i^{th}$ object can then be written as

$$x_i(t) = g_{s_i} \eta_{s_i} + \sqrt{1 - g_{s_i}^2} \epsilon_i \tag{2}$$

where the cluster-related influences are driven by $\eta_{s_i}$ and the object-specific effects by $\epsilon_i$, both treated as Gaussian random variables with unit variance and zero mean[1]. The relative contribution is controlled by the intra-cluster coupling parameter $g_{s_i}$. The Noh-Giada-Marsili model encodes the idea that objects which have something in common belong in the same cluster, object membership in a particular cluster is mutually exclusive and intra-cluster correlations are positive.

If one takes Equation 2 as a statistical hypothesis, it is possible to compute the probability density $P\left(\{\bar{x_i}\}|\mathcal{G},\mathcal{S}\right)$ for any given set of parameters $(\mathcal{G},\mathcal{S}) = (\{g_s\},\{s_i\})$ by observing the data set $\{x_i\}, i, s = 1, ..., N$ as a realisation of the common component of Equation 2 as follows Giada and Marsili (2001):

$$P\left(\{\bar{x_i}\}|\mathcal{G},\mathcal{S}\right) = \prod_{d=1}^{D} \left\langle \prod_{i=1}^{N} \delta\left(x_i(t) - (g_{s_i}\bar{\eta}_{s_i} + \sqrt{1-g_{s_i}^2}\bar{\epsilon}_i)\right)\right\rangle. \tag{5}$$

In Equation 5, $N$ is the number of objects and $D$ is the number of feature measurements for each object. The variable $\delta$ is the Dirac delta function and $\langle ... \rangle$ denotes the mathematical expectation. For a given cluster structure $\mathcal{S}$, the likelihood is maximal when the parameter $g_s$ takes the values

$$g_s^* = \begin{cases} \sqrt{\frac{c_s - n_s}{n_s^2 - n_s}} & \text{for } n_s > 1, \\ 0 & \text{for } n_s \leq 1. \end{cases} \tag{6}$$

$n_s$ in Equation 6 denotes the number of objects in cluster $s$, i.e.

$$n_s = \sum_{i=1}^{N} \delta_{s_i,s}. \tag{7}$$

The variable $c_s$ is the internal correlation of the $s^{th}$ cluster, denoted by the following equation:

$$c_s = \sum_{i=1}^{N}\sum_{j=1}^{N} C_{ij}\delta_{s_i,s}\delta_{s_j,s}. \tag{8}$$

The variable $C_{ij}$ is the *Pearson correlation coefficient* of the data, denoted by the following equation:

$$C_{ij} = \frac{\bar{x_i}\bar{x_j}}{\sqrt{\|\bar{x_i}^2\|\|\bar{x_j}^2\|}}. \tag{9}$$

The maximum likelihood of structure $\mathcal{S}$ can be written as $P\left(\mathcal{G}^*,\mathcal{S}|\bar{x_i}\right) \propto \exp^{D\mathcal{L}(\mathcal{S})}$, where the resulting likelihood function per feature $\mathcal{L}_c$ is denoted by

$$\mathcal{L}_c(\mathcal{S}) = \frac{1}{2}\sum_{s:n_s>1}\left(\log\frac{n_s}{c_s} + (n_s - 1)\log\frac{n_s^2 - n_s}{n_s^2 - c_s}\right). \tag{10}$$

[1]This form of the price model ensures that the self correlation of a stock is one and independent of the cluster coupling. This can be seen by computing the self correlation $E[x_i^2]$ and using that clusters and stock unique process are unit variance zero mean processes

$$\mathrm{E}[(g_{s_i}\eta_{s_i} + \sqrt{1-g_{s_i}^2}\epsilon_i)^2] = g_{s_i}^2 + (1 - g_{s_i}^2) = 1. \tag{3}$$

This is not a unique choice, another possible choice often used is

$$\mathrm{E}[(\frac{\sqrt{g_{s_i}}}{\sqrt{1+g_{s_i}}}\eta_{s_i} + \frac{1}{\sqrt{1+g_{s_i}}}\epsilon_i)^2] = \frac{1+g_{s_i}}{1+g_{s_i}} = 1. \tag{4}$$

From Equation 10, it follows that $\mathcal{L}_c = 0$ for clusters of objects that are uncorrelated, i.e. where $g_s^* = 0$ or $c_s = n_s$ or when the objects are grouped in singleton clusters for all the cluster indexes ($n_s = 1$). Equations 8 and 10 illustrate that the resulting maximum likelihood function for $\mathcal{S}$ depends on the *Pearson correlation coefficient* $C_{ij}$ and hence exhibits the following advantages in comparison to conventional clustering methods:

- It is **unsupervised**: The optimal number of clusters is unknown *a priori* and not fixed at the outset
- The interpretation of results is **transparent** in terms of the model, namely Equation 2.

Giada and Marsili state that $\max_s \mathcal{L}_c(\mathcal{S})$ provides a measure of structure inherent in the cluster configuration represented by the set $\mathcal{S} = \{s_1, ..., s_n\}$ Giada and Marsili (2001). The higher the value, the more pronounced the structure.

We note that the particular choice of Gaussian innovations in Equation 2 is convenient, since the *Pearson correlation coefficient* then completely characterises pairwise interactions amongst objects in the system (Giada and Marsili (2001)). This is a necessary condition, given the physical analogy and link to the motivating Hamiltonian given in Equation 1. The application of this technique to high-frequency financial time series may motivate a more prudent assumption for the underlying object and cluster dynamics, incorporating jumps to better model the price formation process at this scale. However, the use of, say, jump diffusion innovations would require an alternative dependency metric, such as Lévy copulas, to completely capture object interactions (Cont and Tankov (2004), McNeil et al. (2015)), requiring a careful re-derivation of the appropriate likelihood function. This will be explored in further research.

## 4. Detecting temporal states using clustering

The data generative model specified by Equation 2 is sufficiently generic that it can be applied to a diverse set of problem domains, where object and cluster innovations can be assumed to be Gaussian. In the financial domain, initial applications focused on clustering stocks based on price changes (Giada and Marsili (2001), Hendricks et al. (2015)), however Marsili (2002) proposed that this technique could be used to cluster *time periods* in order to identify *temporal market states*. Days were grouped into clusters based on the closing price performance of the chosen universe of stocks, demonstrating a meaningful classification of market-wide activity which persists through time (Marsili (2002)). We propose that a similar approach can be applied to discover *intraday* temporal states, clustering time periods based on the performance of *multiple* observable market microstructure features. A practical trading system often has access to a real-time market data feed, from which multiple features can be extracted to describe various aspects of the evolving limit order book. In addition, examining temporal cluster configurations at varying time scales can suggest a hierarchy of system behaviour, providing insights into exogenous and endogenous market activity. This can also assist trading agents in developing optimal trajectories for varying objectives, such as stock acquisition or liquidation at minimal cost. In particular, for an agent tasked to learn an optimal policy (state-action mapping), the grouping of temporal periods into market states based on market microstructure feature performance provides a novel scheme to reduce the dimensionality of the state space and promote efficient learning.

In this paper, we will focus on the emergent hierarchy of system behaviour at different time scales and explore a scheme for online state detection. In one application, this leads to a system of 1-step state transition probability matrices at varying scales, which can be refined online in real-time. These can be used in optimal planning schemes where Markovian dynamics are assumed and state persistence can be exploited.

## 5. A high-speed Parallel Genetic Algorithm implementation

The likelihood function specified in Equation 10 serves as the objective function in a metaheuristic optimisation routine, where candidate cluster configurations are evaluated and successively improved until a configuration best explains the inherent structure in a given correlation matrix. Giada and Marsili (2001) used simulated annealing and deterministic maximisation to approximate the maximum likelihood structure. While appropriate for their study, these techniques are inherently computationally intensive and may require a significant amount of time to converge for large-scale problems. In addition, it is unclear whether such trajectory-based methods are appropriate for the multi-featured clustering problem considered in this paper, since Giada and Marsili (2001) clustered objects (stocks) based on a single feature (price returns). Hendricks et al. (2015) and Cieslakiewicz (2014) propose the use of a high-speed Parallel Genetic Algorithm (PGA), leveraging the Streaming Multiprocessors (SMs) of a Graphics Processing Unit (GPU), where Equation 10 is used as a fitness function to find the cluster configuration which best approximates the maximum likelihood structure. They implemented a C-based master-slave PGA using the Nvidia Compute Unified Device Architecture (CUDA) development environment, using the Single Program Multiple Data (SPMD) architecture to enumerate the GPU thread hierarchy with population members for concurrent application of genetic operators.

Consider the problem of finding the cluster configuration of $n$ objects. Then, given $N$ candidate cluster configuration structures making up the population,

$$\begin{aligned} \mathcal{S}_1 &= \{s_1^1, ..., s_n^1\} \\ \mathcal{S}_2 &= \{s_1^2, ..., s_n^2\} \\ &\vdots \\ \mathcal{S}_N &= \{s_1^N, ..., s_n^N\} \end{aligned}$$

would be mapped to the GPU thread hierarchy using a 2-dimensional grid, as shown in Table 1.

| CUDA thread block grid | | | | |
|---|---|---|---|---|
| | $\mathcal{S}_1$ | $\mathcal{S}_2$ | ... | $\mathcal{S}_N$ |
| $object_1$ | $s_1^1$ | $s_1^2$ | ... | $s_1^N$ |
| $object_2$ | $s_2^1$ | $s_2^2$ | ... | $s_2^N$ |
| $\vdots$ | $\vdots$ | $\vdots$ | ... | $\vdots$ |
| $object_n$ | $s_n^1$ | $s_n^2$ | ... | $s_n^N$ |

**Table 1.:** Mapping of population to CUDA thread hierarchy

The PGA was applied to the relatively small problem of finding the cluster configuration of 18 objects, however demonstrated fast absolute computation time compared to state-of-the-art methods, with the promise of scalability within the constraints of the GPU architecture used (Hendricks et al. (2015), Cieslakiewicz (2014)). We have restricted our analysis to intraday temporal periods within one month, however this still yields up to 2208 objects in the 5-minute case. Table 2 shows the specifications and capabilities of the two candidate GPUs and Table 3 shows the PGA parameter values and number of objects for each of the time scales investigated. The mapping of candidate configurations to the GPU thread hierarchy under the SPMD paradigm results in an upper bound on the permissible number of objects and population size. Hendricks et al. (2015) further recognised the importance of ensuring that the population size is large enough relative to the number of objects, to ensure sufficient population diversity for convergence to the best approximation of the maximum likelihood structure within a finite number of generations. Smaller populations often lead to sub-optimal algorithm terminations and inconsistent results. For the 60-minute, 30-minute and 15-minute cases, the Nvidia Geforce GTX765m notebook GPU had sufficient capability to determine the optimal cluster configurations from sufficiently large populations. The 5-minute case demanded a larger capacity GPU, and the Nvidia Geforce GTX Titan X provided the necessary additional SMs, CUDA cores and global memory to facilitate efficient computation.

| | Graphics Processing Unit (GPU) | |
|---|---|---|
| Feature | Nvidia Geforce GTX 765m | Nvidia Geforce GTX Titan X |
| Compute capability | 3.5 | 5.2 |
| CUDA cores | 768 | 3072 |
| Memory | 2048MB | 12228MB |
| Number of streaming multiprocessors | 16 | 96 |
| Max threads / thread block | 1024 | 1024 |
| Thread block dimension | 32 | 32 |
| Max thread blocks / multiprocessor | 16 | 32 |

**Table 2.:** Graphics Processing Unit specification and capabilities

| Time scale | Number of periods (objects) | Population size | Generations | Stall generations | Mutation probability | Crossover probability | Computation Time (sec)∗ |
|---|---|---|---|---|---|---|---|
| 5-minute | 2208 | 4000 | 4000 | 1000 | 0.09 | 0.9 | 603 ($D$) |
| 15-minute | 736 | 1000 | 4000 | 500 | 0.09 | 0.9 | 382 ($N$) |
| 30-minute | 368 | 800 | 4000 | 500 | 0.09 | 0.9 | 215 ($N$) |
| 60-minute | 184 | 600 | 4000 | 500 | 0.09 | 0.9 | 132 ($N$) |

**Table 3.:** Parameter values and computation times for Parallel Genetic Algorithm
∗ Average from 20 independent runs; $N$ refers to the GTX765m Notebook GPU and $D$ refers to the GTX Titan X Desktop GPU.

We note that the number of generations and stall generations indicated in Table 3 are higher than one would typically specify for a genetic algorithm, since these promote potential over-fitting to the prescribed dataset. Recall that our application is to find the candidate cluster configuration which best explains the structure inherent in a given correlation matrix. Thus we are not concerned with out-of-sample validity, but would rather prefer to find a configuration with the highest likelihood value. The higher number of generations and stall generations, together with the mutation operator, promotes convergence to a higher likelihood structure. The average computation times indicated in Table 3 are not overly onerous, suggesting that for practical application, overnight or even intraday estimation of cluster configurations to capture recent dynamics is feasible. The proposed PGA thus offers an efficient, scalable alternative for finding the best approximation of the optimal cluster configuration, suitable for clustering objects on multiple observable features.

## 6. State Signature Vectors for online state detection

The clustering procedure described thus far can be used as an unsupervised algorithm to group temporal periods into states according to feature similarity, however this can only reveal the *ex-ante* temporal states and is not suitable for online detection. Upon examination of the resulting cluster configurations, we noted that each node refers to a particular time period, with an associated signature of market activity. Furthermore, if two time periods appear in the same cluster, given the data generative model assumed in Equation 2, we conjecture that it is the relative similarity of their characteristic signatures of market activity which resulted in their assignment to the same cluster. Using this idea, given a cluster configuration of temporal periods into market states, it is possible to extract a *state signature vector* (SSV) which summarises the signature of market activity across stocks and time periods for each state. Then, if one is faced with a new candidate feature vector (FV), the market state assignment can be determined by using the closest match within the set of pre-determined SSVs computed offline. FVs are easy to compute online from a streaming datafeed and state assignment can be achieved using a simple Euclidean distance computation. To make these ideas concrete, consider the example illustrated in Figure 1.

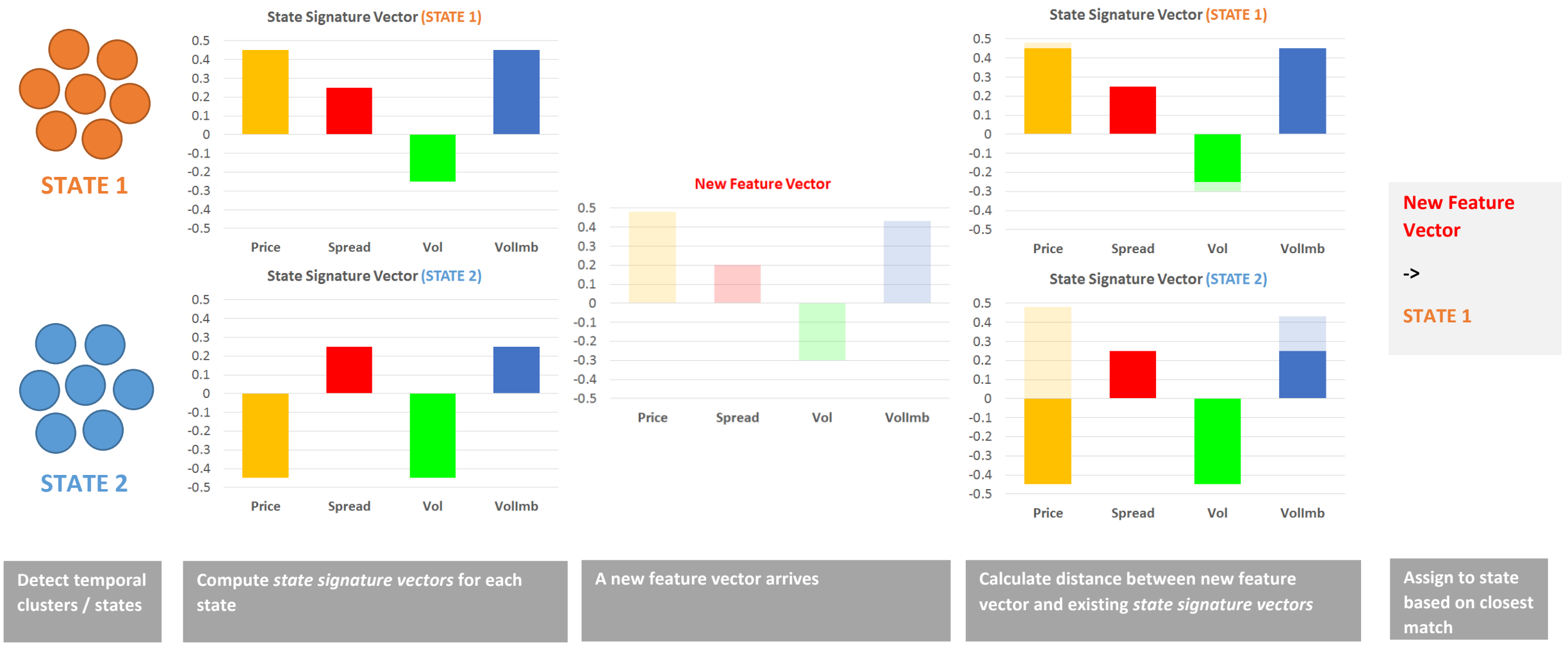

**Figure 1.:** Illustration of online state assignment based on identified state signature vectors.

Here, we compute two SSVs from the identified states, and use these as a basis for assigning a new FV to a market state. This is based on a simple Euclidean distance metric,

$$\text{argmin}_p ||FV - SSV_p||,$$

where $p$ is the index of the identified states.

In this paper, we have used four features to characterise market activity at intraday scale. These include: *trade price, trade volume, spread* and *quote volume imbalance*. In particular, we consider the *relative change* in each of these features. For example, based on a set of feature measurements $\mathcal{F}^{5min}$ at 5-minute scale, we would compute

$$\triangle f_t^{5min} = \frac{f_t^{5min} - f_{t-1}^{5min}}{f_{t-1}^{5min}}$$

for all $f_t^{5min} \in \mathcal{F}^{5min}$. For the initial temporal cluster detection stage, these "feature returns" are calculated for each stock and concatenated before computing the time period correlation matrix.

For the extraction of SSVs from significant states, we compute average feature returns across member periods and stocks. For example, consider the case of 15-minute period clustering. If one state (cluster) consisted of 2 periods (09:15 - 09:30 and 15:15 - 15:30), then we would find the average *trade price, trade volume, spread* and *quote volume imbalance* returns across stocks in each period (i.e. two 4-element vectors), then average across these two vectors to get a single 4-element vector, which would be the representative SSV for that state.

Although this results in a loss of information, we conjecture that the average signature of feature returns broadly captures the state of market activity. The SSVs for each time-scale configuration are illustrated in Figures 4, 6, 8 and 10. Following this approach, the FVs calculated in the online environment would constitute the same averages of feature returns, before matching to the appropriate SSV. We note that this is merely one candidate scheme for extracting SSVs which are conducive to online matching for state assignment, however alternative schemes for extraction of SSVs which preserve state-specific information will be explored in future work. The chosen features do not represent an exhaustive set of possible explanatory factors for intraday market activity, but rather were chosen based on the relative ease of their online construction from streaming Level-1 market data feeds JSE (2015). Additional features can be considered in future work.

## 7. Scale-invariant characteristics of states

The detected temporal cluster configurations can be further analysed to determine whether any characteristics exhibit scale-invariant behaviour. In particular, a visual inspection of the cluster configurations shown in Section 8.4 led us to conjecture a possible power-law fit for cluster sizes. Many physical and man-made systems exhibit characteristics which follow a power-law functional form, and its unique mathematical properties sometimes lead to surprising physical insights (Gabaix et al. (2003), Clauset et al. (2009)). Many authors have investigated the nature of information and forecasting at different time scales in financial markets (see Dacorogna et al. (1996), Zhang et al. (2005), Emmert-Streib and Dehmer (2010) as examples). For our application, the existence of different critical exponents for the best power-law fits at different time scales may suggest different universality classes which characterise the system activity at each scale. In fact, Mastromatteo and Marsili (2011) discuss the notion that, for a complex adaptive system, distinguishable models can only be gleaned when the system is near criticality. Thus, if financial markets truly are a complex adaptive system, measurable quantities from the dynamics at each scale should yield a statistically significant power-law fit. Although it is difficult to quantify the exact nature of these scale-specific behaviours or universality classes, their apparent existence suggests that investment and trading decisions would benefit from time-scale-specific state space information. This would enhance the efficacy of intraday policies which aim to find optimal trajectories through the system.

Given the difficulties of identifying statistically significant power-law fits to empirical quantities Bauke (2007), we incorporated the maximum likelihood fitting procedure provided by Clauset et al. (2009). Outputs from their functions include the scaling parameter of the proposed power-law fit, a Kolmogorov-Smirnov test for the goodness-of-fit of the proposed model to the data, the lower-bound for the fit if the tail distribution follows a power-law and the log-likelihood of the data under the power-law fit.

We note that a detected temporal cluster configuration results in a set of homogeneous market states, although it is not clear which states are significant, i.e. likely to persist, or merely transient. Using *all* identified states may result in spurious state assignments if one uses the online algorithm described in Section 6. This leads to the need for some selection criteria for significant states, before extracting SSVs. Candidate criteria include using intra-cluster connectedness ($c_s$) or cluster size with some form of thresholding procedure, however these heuristics are inherently subjective. The power-law fit to cluster size provides one candidate objective approach for state selection. Under the assumption that the system is near criticality when we find a stable parameter calibration, choosing the states which best fit the power-law functional form may aid in isolating those states which best capture the system behaviour at that scale, i.e. filter the stable, persistent states from the noise. This provides an objective mechanism for selecting significant states, reducing the set of SSVs which form the basis for the online state detection algorithm.

## 8. Data and results

### 8.1. *Data description*

The data for this study constituted tick-level trades and top-of-book quotes for 42 stocks on the Johannesburg Stock Exchange (JSE) from 1 November 2012 to 30 November 2012. This data was sourced from the Thomson Reuters Tick History (TRTH) database. The raw data was aggregated according to the time-scale considered (5-minute, 15-minute, 30-minute and 60-minute), before calculating the required features (change in trade price, trade volume, spread and volume imbalance).

The 42 stocks considered represent the prevailing constituents of the FTSE/JSE Top40 headline index, which contains the 42 largest stocks by market capitalisation in the main board's FTSE/JSE All-Share index.

The objects of interest for the cluster analysis are the *time periods*. Table 4 provides an example of the required data returns matrix, from which a correlation matrix is computed for time period similarity. This is the only required input for the clustering algorithm.

| | **Feature** | **Times** | | | | | |
|---|---|---|---|---|---|---|---|
| | | 01-Nov-2012 09:00 | 01-Nov-2012 09:15 | 01-Nov-2012 09:30 | ... | 30-Nov-2012 16:30 | 30-Nov-2012 16:45 |
| **Trade Price** | AGL trade price return | 0.35 | 0.60 | 0.85 | ... | 0.39 | 0.22 |
| | AMS trade price return | 0.94 | 0.71 | 0.73 | ... | 0.63 | 0.78 |
| | SBK trade price return | 0.70 | 0.38 | 0.58 | ... | 0.38 | 0.81 |
| | ⋮ | ⋮ | ⋮ | ⋮ | ⋮ | ⋮ | ⋮ |
| | WHL trade price return | 0.90 | 0.49 | 0.05 | ... | 0.65 | 0.53 |
| **Spread** | AGL spread return | 0.64 | 0.49 | 0.68 | ... | 0.05 | 0.95 |
| | AMS spread return | 0.33 | 0.09 | 0.76 | ... | 0.44 | 0.97 |
| | SBK spread return | 0.09 | 0.73 | 0.54 | ... | 0.80 | 0.48 |
| | ⋮ | ⋮ | ⋮ | ⋮ | ⋮ | ⋮ | ⋮ |
| | WHL spread return | 0.41 | 0.61 | 0.11 | ... | 0.40 | 0.69 |
| **Trade Volume** | AGL trade volume return | 0.61 | 0.59 | 0.96 | ... | 0.65 | 0.50 |
| | AMS trade volume return | 0.16 | 0.09 | 0.47 | ... | 0.86 | 0.57 |
| | SBK trade volume return | 0.98 | 0.05 | 0.67 | ... | 0.72 | 0.12 |
| | ⋮ | ⋮ | ⋮ | ⋮ | ⋮ | ⋮ | ⋮ |
| | WHL trade volume return | 0.38 | 0.49 | 0.36 | ... | 0.27 | 0.81 |
| **Volume Imbalance** | AGL volume imb return | 0.01 | 0.45 | 0.78 | ... | 0.69 | 0.77 |
| | AMS volume imb return | 0.54 | 0.17 | 0.87 | ... | 0.47 | 0.44 |
| | SBK volume imb return | 0.20 | 0.42 | 0.91 | ... | 0.88 | 0.58 |
| | ⋮ | ⋮ | ⋮ | ⋮ | ⋮ | ⋮ | ⋮ |
| | WHL volume imb return | 0.20 | 0.09 | 0.38 | ... | 0.90 | 0.12 |

**Table 4.:** Illustration of data returns matrix as an input for estimation of 15-minute period correlations

## 8.2. *Workflow*

Figure 2 illustrates the process workflow and tools used for performing the temporal cluster analysis. The TRTH tick data is stored in a MongoDB noSQL database, with optimised query indexes for efficient data retrieval. A bespoke Application Programming Interface (API) was written to transport data from MongoDB to our primary scientific computing platform, MATLAB. The data is used to instantiate a High Frequency Time Series (HFTS) object in MATLAB, which allows for efficient merging, resampling and aggregation of large-scale irregularly-spaced tick data. Based on a chosen time-scale, the data is aggregated, features are extracted and returns calculated, before computing the time period correlation matrix. The PGA was implemented in CUDA-C using Nvidia Nsight and the Microsoft Visual Studio development environment. The compiled PGA was called from the MATLAB environment to run the temporal cluster analysis. The resulting cluster configuration is transported to the MATLAB workspace, from which we can determine the power-law fits, extract SSVs, estimate online clusters and compute transition probability matrices. Using the stock, time period, cluster configuration and correlation data, a MATLAB script was written to generate an XML file containing the required node and edge metadata for an undirected graph to import into Gephi. Gephi was used for cluster configuration visualisation, as described in Section 8.3.

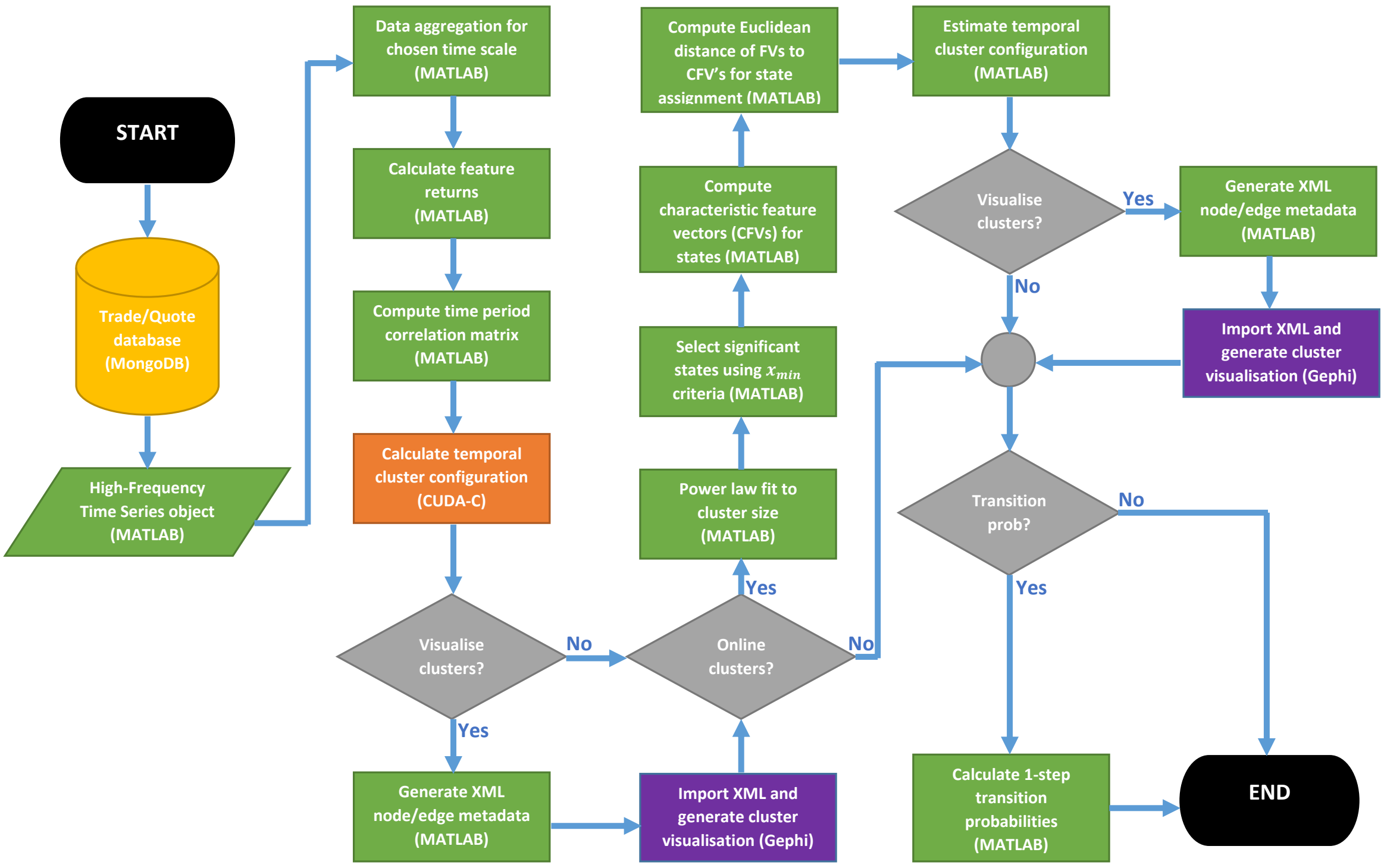

**Figure 2.:** Flowchart illustrating workflow to determine the temporal cluster configuration from a time period correlation matrix, identify persistent states, estimate temporal cluster configuration using feature vectors and determine state transition probabilities. Processes are coloured by platform: MongoDB = Yellow, MATLAB = Green, CUDA-C = Orange, Gephi = Purple.

### 8.3. *Visualisation*

For the cluster configuration visualisation, we made use of the Gephi graph visualisation and manipulation software package (Bastian et al. (2009)), with a customised enumeration of nodes and edges and the Fruchterman and Reingold (1991) node spacing algorithm. The presence of an edge between nodes indicates membership to the same cluster, while edge thickness provides a visual impression of object-object correlation, and hence intra-cluster connectedness. For the visualisations which follow, we chose to colour the nodes by intraday time period, in order to illuminate any calendar time effects in the detected states. According to this scheme, the same time on different days will receive the same colour. These visualisations are shown in Figures 3, 5, 7, 9, 12, 13, 14 and 15.

### 8.4. *Results discussion*

For each set of results, we consider 8 hours of continuous trading activity each day, from 09:00 to 17:00, for the duration of one month. Figure 3 shows the temporal cluster configuration of 60-minute periods. We first note that the detection of non-trivial clusters from microstructure-based time correlations indicates that intraday dynamics may be reducible to a finite set of temporal states. Considering the time-of-day colour shading, we notice two clusters which exhibit market activity characteristics which coincide with morning and afternoon times. The dark green cluster refers to the first hour of the trading day (09:00 to 10:00), which incorporates opening auction and subsequent activity. We note that the South African equity market is strongly influenced by global market activity, in part due to local stocks being listed on multiple exchanges in the UK, USA, Europe and Australia (JSE (2014)). During the period considered in this analysis, the UK market open occurred at 10:00 SAST and US market open at 15:30 SAST. The UK market open has a significant impact on local trading dynamics, with the 10:00 to 11:00 periods dispersing across clusters with no discernible time-of-day correlation. We note a contiguous dark orange cluster emerge from 15:00 to 16:00, as the US market starts to participate in local trading activity. This pattern of market activity broadly corroborates these exogenous market effects from global markets. Figure 4 shows the SSVs extracted from the significant states selected from Figure 3. As discussed in Section 7, we used the $x_{min}$ statistic from the power-law fit to the tail distribution of cluster sizes to determine the significant states. For the 60-minute periods, the most significant power-law fit was for cluster sizes $\geq 13$, resulting in 6 significant states. The resulting SSVs are all relatively different, when considering the magnitude and direction of each of the average change in feature values. This ensures greater certainty in the state assignment of an online FV.

Figure 5 shows the temporal cluster configuration of 30-minute periods. We see a larger number of states emerge as the granularity increases, with 60-minute states being dissected based on finer-grained market activity. The dark green and dark orange contiguous morning and afternoon states still persist at this scale, although endogenous system characteristics begin to mask previously identified exogenous characteristics. We note that there is no defined hierarchy emerging, in that a set of 30-minute clusters cannot be combined to form the 60-minute clusters identified previously, further highlighting time-scale-specific behaviour. Figure 6 shows the SSVs of significant states, based on the 10 clusters with a size $\geq 14$.

Figure 7 shows the temporal cluster configuration of 15-minute periods. We notice increasing time-of-day diversity in each of the identified clusters, further highlighting endogenous system activity. The red contiguous cluster is associated with the period from 16:30 to 16:45, suggesting a particular signature of market activity leading into the closing auction, which starts at 16:50. The UK and US related effects seem to have a weaker impact at this scale, with exchange-specific rules having a more dominant effect. As a result, we see a larger variety of SSVs in Figure 8, some with similar profiles seen at the 30-minute scale, but with a larger focus on magnitude, rather than merely direction.

Figure 9 shows the temporal cluster configuration of 5-minute periods. Here we see quite a different profile of system behaviour. There are a large number of singletons, which could be attributed to the amount of noise in the data at this scale, making it more difficult to discern significant structure. We notice an interesting time-of-day correlation with detected clusters, however broad periods (morning, lunch, afternoon) appear to have been dissected into contiguous blocks based on state-specific market activity. The 5-minute time scale is starting to capture the effects of automated, rule-based trading agents which shows quite a different characteristic signature. This further highlights the importance of studying market activity profiles at the scale at which you intend to participate. Even when one considers the associated SSVs in Figures 10 and 8, the 5-minute and 15-minute studies exhibit the same number of significant states using the power-law criterion, however the combinations of direction and magnitude for the feature values are quite different.

Figure 11 illustrates the results of the power-law fits to the cluster size empirical distribution at each time scale. Each fit to the tail distribution exhibits a Kolmogorov-Smirnov $p$-value $> 0.1$ (assuming a null hypothesis of a power-law fit), suggesting a strong fit of the power-law functional form for the given scaling factor ($\alpha$) and minimum size ($x_{min}$) (Clauset et al. (2009)). In addition, we note the $\alpha$ exponents are different for each of the time scales considered. This evidence of statistically significant power-law fits at each measured scale is consistent with the notion of financial markets as a complex adaptive system, and that the system is near criticality at each measured time scale (Mastromatteo and Marsili (2011)). A further study should verify whether this suggests different universality classes of system behaviour at different time scales, however these preliminary results do indicate the presence of some complex hierarchy of system behaviour, motivating the need for scale-specific temporal analysis.

Figure 12 shows the estimated 60-minute cluster configuration for the same period (1 November 2012 to 30 November 2012), but where the distance of each period's FV to the identified SSVs is used as the criterion for state assignment. This is a simple in-sample test to determine whether the proposed scheme for online state assignment can discern the structure suggested by direct application of the clustering algorithm. By comparing Figure 12 and Figure 3, we notice that the online state assignment algorithm does recover the contiguous morning and afternoon states, but more broadly intuitively separates periods into: opening auction and early morning trading state, UK market open state, two lunch states, US market open state and a end-of-day/closing auction state. This completely captures the exogenous market effects, which is a strong validation for the approach. Table 5 shows an empirical 1-step transition probability matrix calculated from the states shown in Figure 12, illustrating one potential application of this technique. The 1-step transitions show a particular preference, suggesting some predictability which can be exploited by trading agents. To be clear, the online assignment of a FV to a state means that we have developed a mechanism to *detect* which state we are currently in, using the prevailing set of SSVs. The transition matrix can be used and updated online, and for optimal planning in the domain.

Figures 13 to 15 and Tables 6 to 8 show the estimated cluster configurations and transition probability matrices using the SSVs at the specified time scale. It is interesting to observe the dilution of the exogenous time-of-day effects as one approaches the 5-minute scale.

Figure 16 illustrates the stability of the online state assignment algorithm out-of-sample. Given that the state assignment of an online FV is based on the minimum Euclidean distance to predetermined SSVs, we compute the *best match* distance for each of the FVs in a sample and use a boxplot to visualise the empirical distribution. This paper proposes offline estimation of SSVs used for online state detection. The online cluster configurations shown in Figures 12, 13, 14 and 15 use FVs from the *ex-ante* period, i.e. the same period used to estimate the SSVs. It is prudent to determine whether state assignment using out-of-sample (*ex-post*) FVs deviate significantly from in-sample assignment, and gauge the out-of-sample efficacy of the SSVs before re-estimation is necessary. Given the computation times shown in Section 5, in practice one could estimate the SSVs overnight for each trading day. We have considered SSVs estimated from the period 1 November 2012 to 30 November 2012, and compared the resulting online states from the *ex-ante* period (1 November 2012 to 30 November 2012) with states from an *ex-post* period (3 December 2012 to 7 December 2012, one week after SSV estimation). From these results, it appears that 60-minute states cannot be reliably determined *ex-post* using the online detection algorithm, given the observed higher range of *best match* Euclidean distances. The 30-minute, 15-minute and 5-minute time scales all exhibit acceptable *ex-post best match* distances, with the exception of a few outliers. From these preliminary results, it appears that the algorithm can be used to reliably determine 30-minute, 15-minute and 5-minute states for a relatively short *ex-post* period following SSV estimation. A more robust study should consider the precise half-life of the SSVs, but given the relatively fast computation time, this is unlikely to be a practical concern.

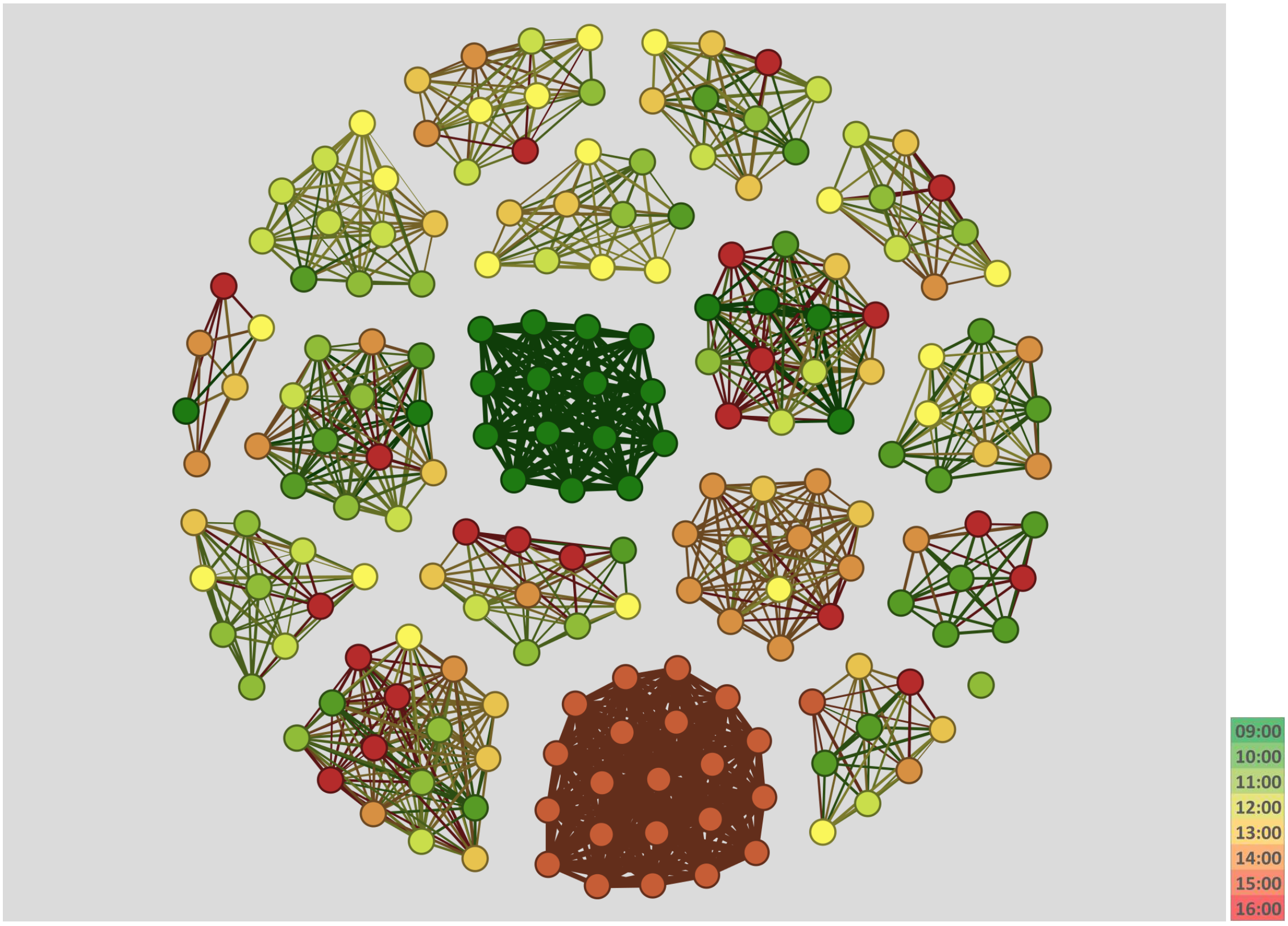

**Figure 3.:** JSE TOP40 60-minute temporal clusters for the period 01-Nov-2012 to 30-Nov-2012, representing 184 distinct periods. Each node represents a 60-minute period during a trading day, with the colour shading indicating the time-of-day (Morning = green, Lunch = yellow, Afternoon = red) and node connectedness indicating cluster membership.

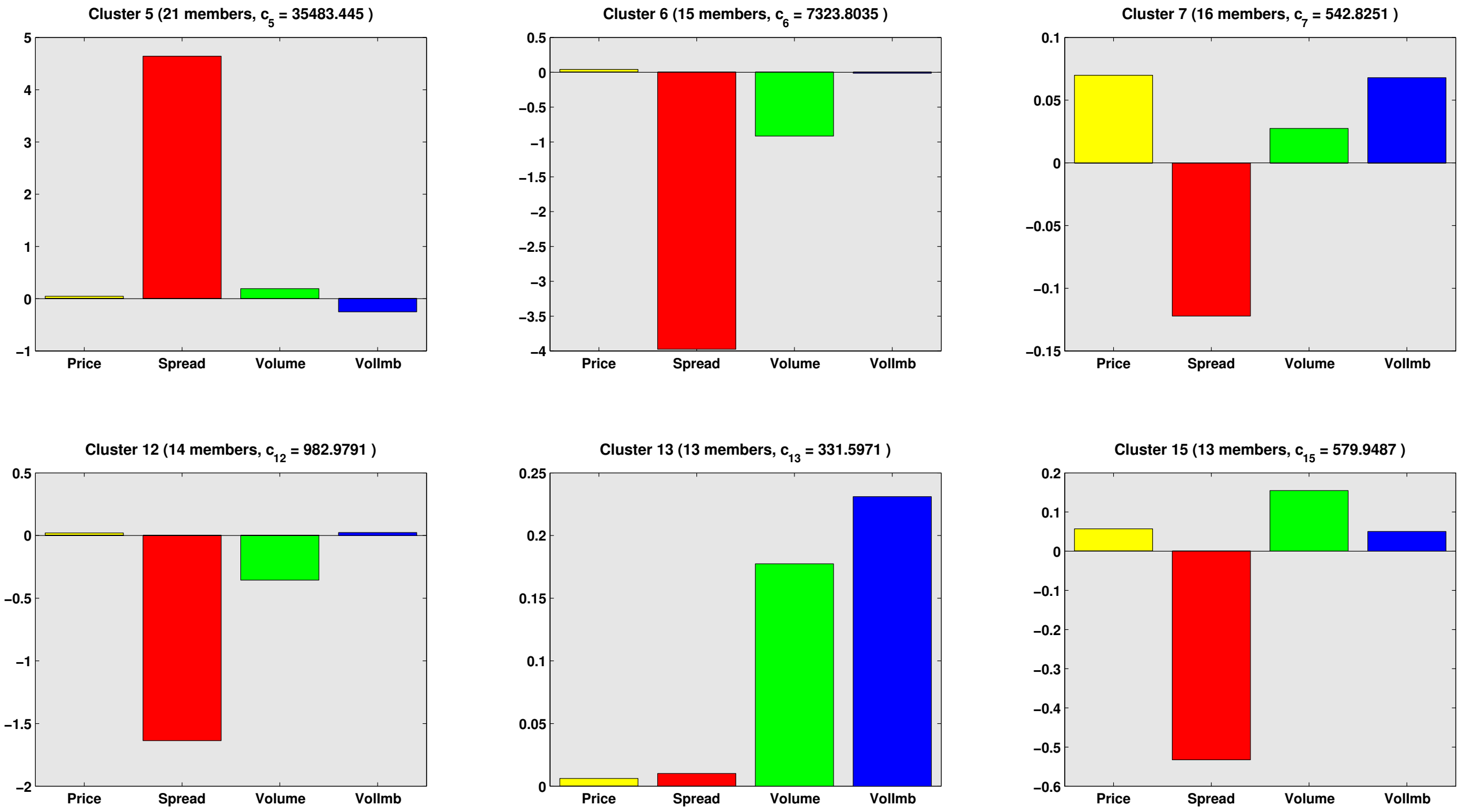

**Figure 4.:** JSE TOP40 60-minute cluster state signature vectors for the period 01-Nov-2012 to 30-Nov-2012. Each plot illustrates the average change in trade price, spread, trade volume and quote volume imbalance across member periods and stocks for each of the clusters with a size $\geq x_{min}$ from the truncated power-law fit. Cluster size and intra-cluster correlation are shown in parentheses.

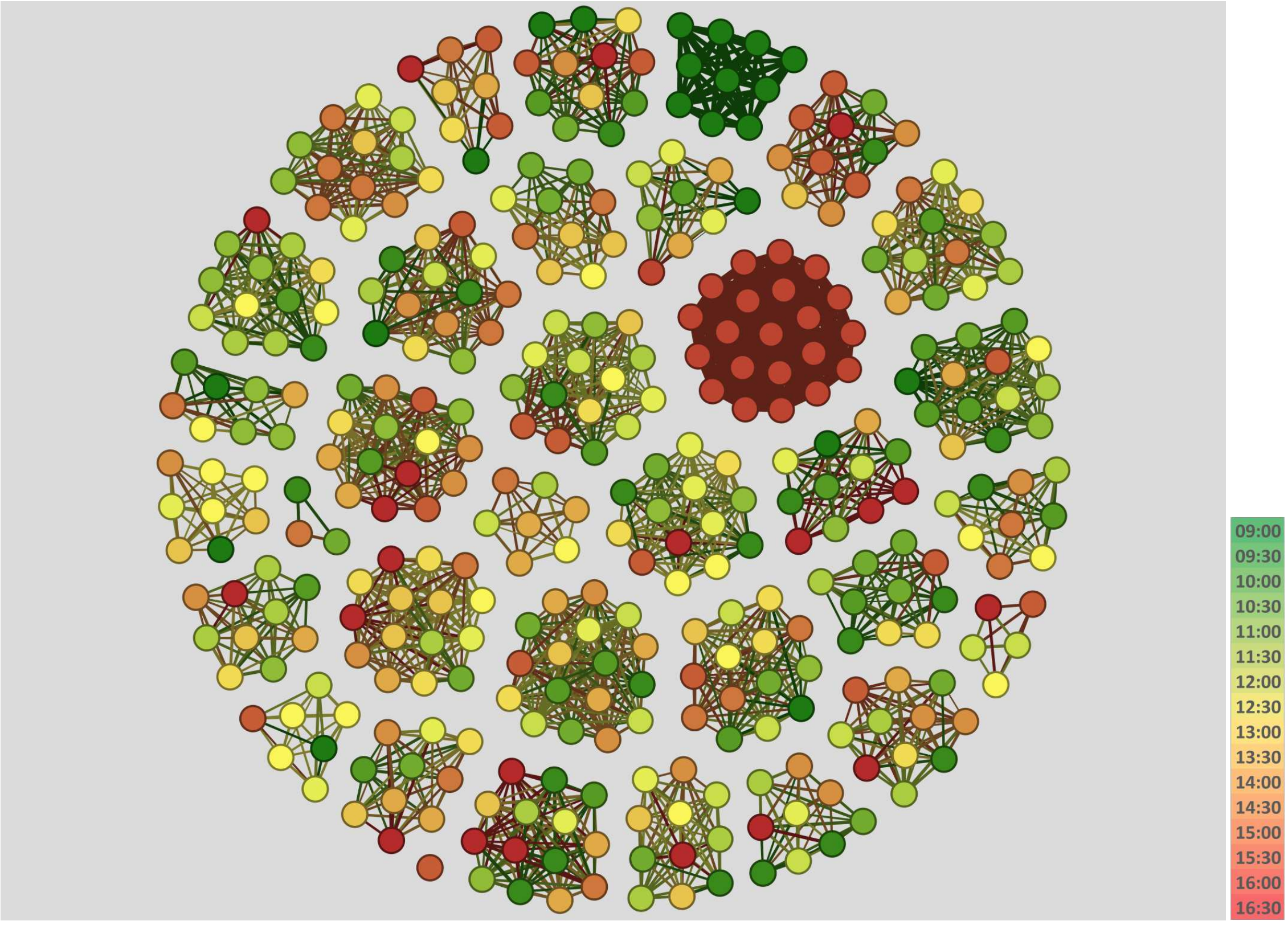

**Figure 5.:** JSE TOP40 30-minute temporal clusters for the period 01-Nov-2012 to 30-Nov-2012, representing 368 distinct periods. Each node represents a 30-minute period during a trading day, with the colour shading indicating the time-of-day (Morning = green, Lunch = yellow, Afternoon = red) and node connectedness indicating cluster membership.

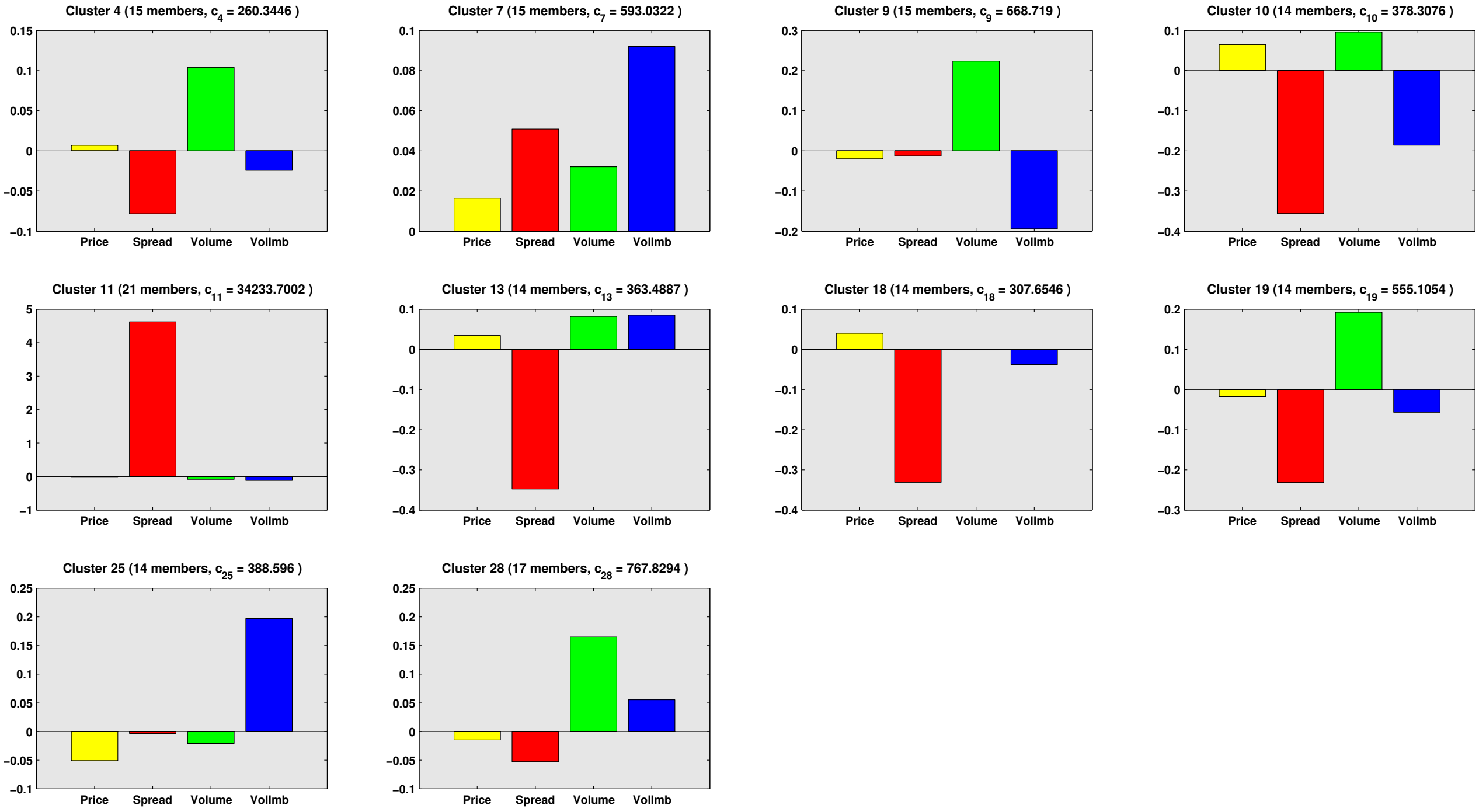

**Figure 6.:** JSE TOP40 30-minute cluster state signature vectors for the period 01-Nov-2012 to 30-Nov-2012. Each plot illustrates the average change in trade price, spread, trade volume and quote volume imbalance across member periods and stocks for each of the clusters with a size $\geq x_{min}$ from the truncated power-law fit. Cluster size and intra-cluster correlation are shown in parentheses.

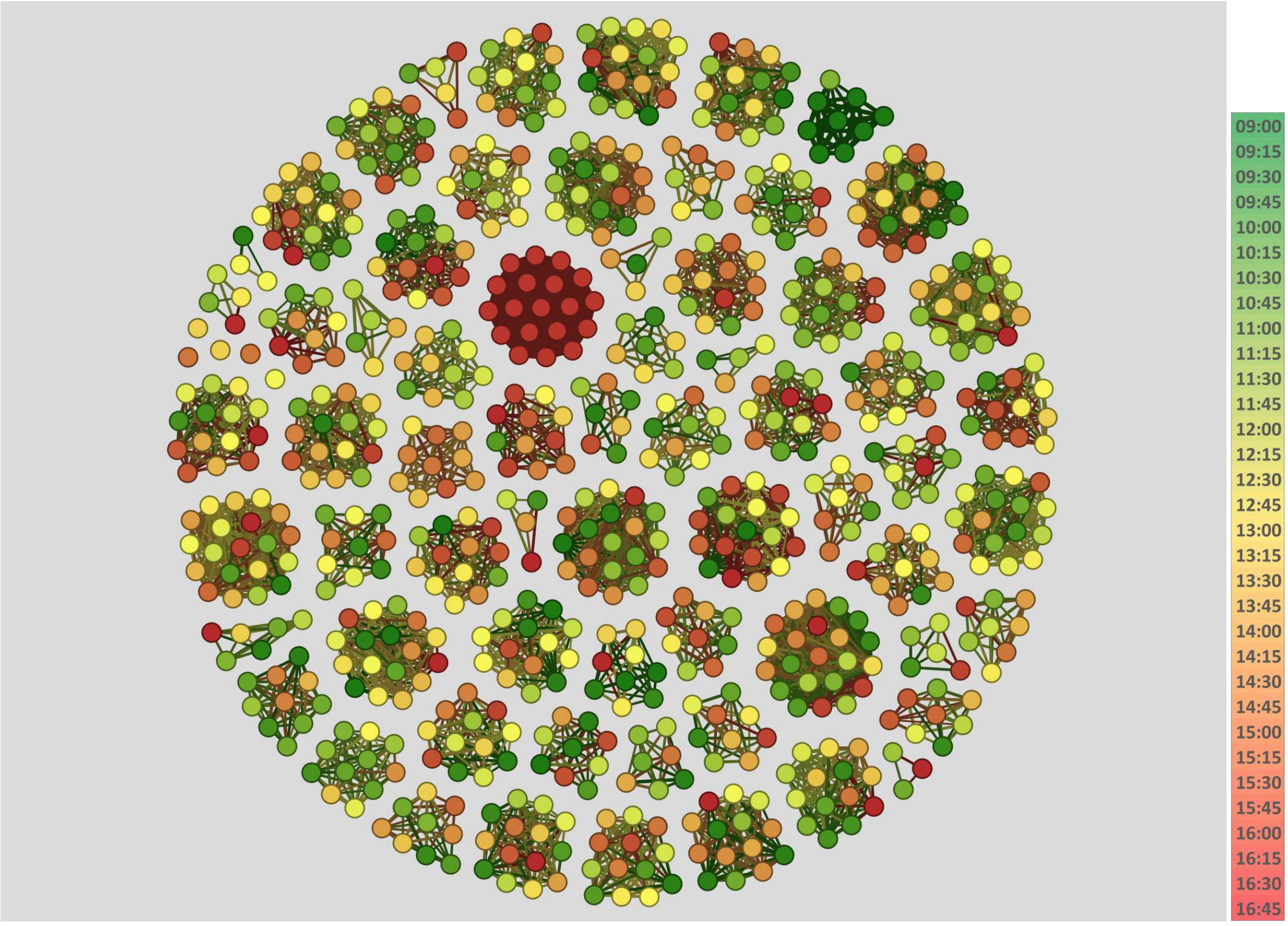

**Figure 7.:** JSE TOP40 15-minute temporal clusters for the period 01-Nov-2012 to 30-Nov-2012, representing 736 distinct periods. Each node represents a 15-minute period during a trading day, with the colour shading indicating the time-of-day (Morning = green, Lunch = yellow, Afternoon = red) and node connectedness indicating cluster membership.

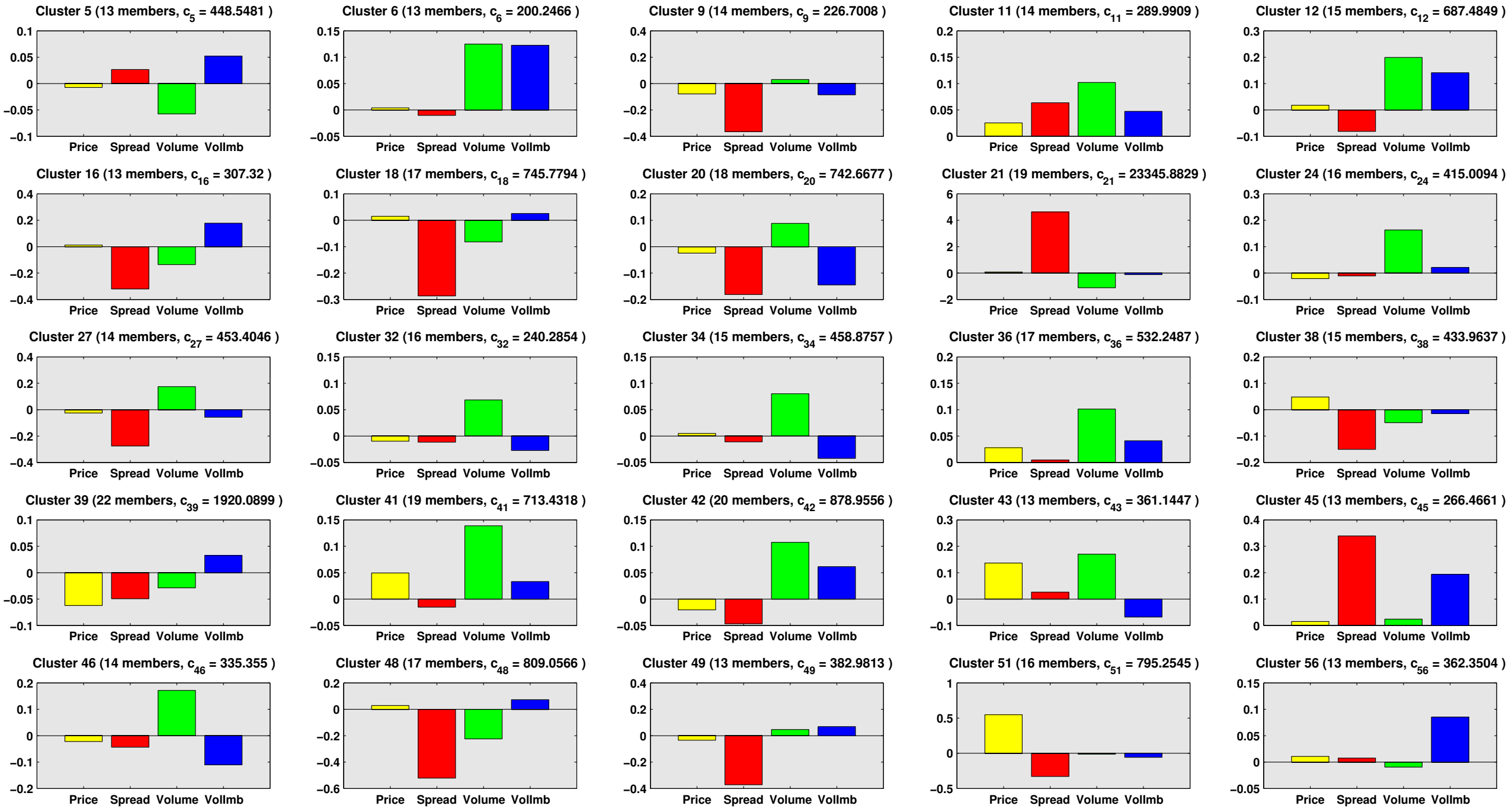

**Figure 8.:** JSE TOP40 15-minute cluster state signature vectors for the period 01-Nov-2012 to 30-Nov-2012. Each plot illustrates the average change in trade price, spread, trade volume and quote volume imbalance across member periods and stocks for each of the clusters with a size $\geq x_{min}$ from the truncated power-law fit. Cluster size and intra-cluster correlation are shown in parentheses.

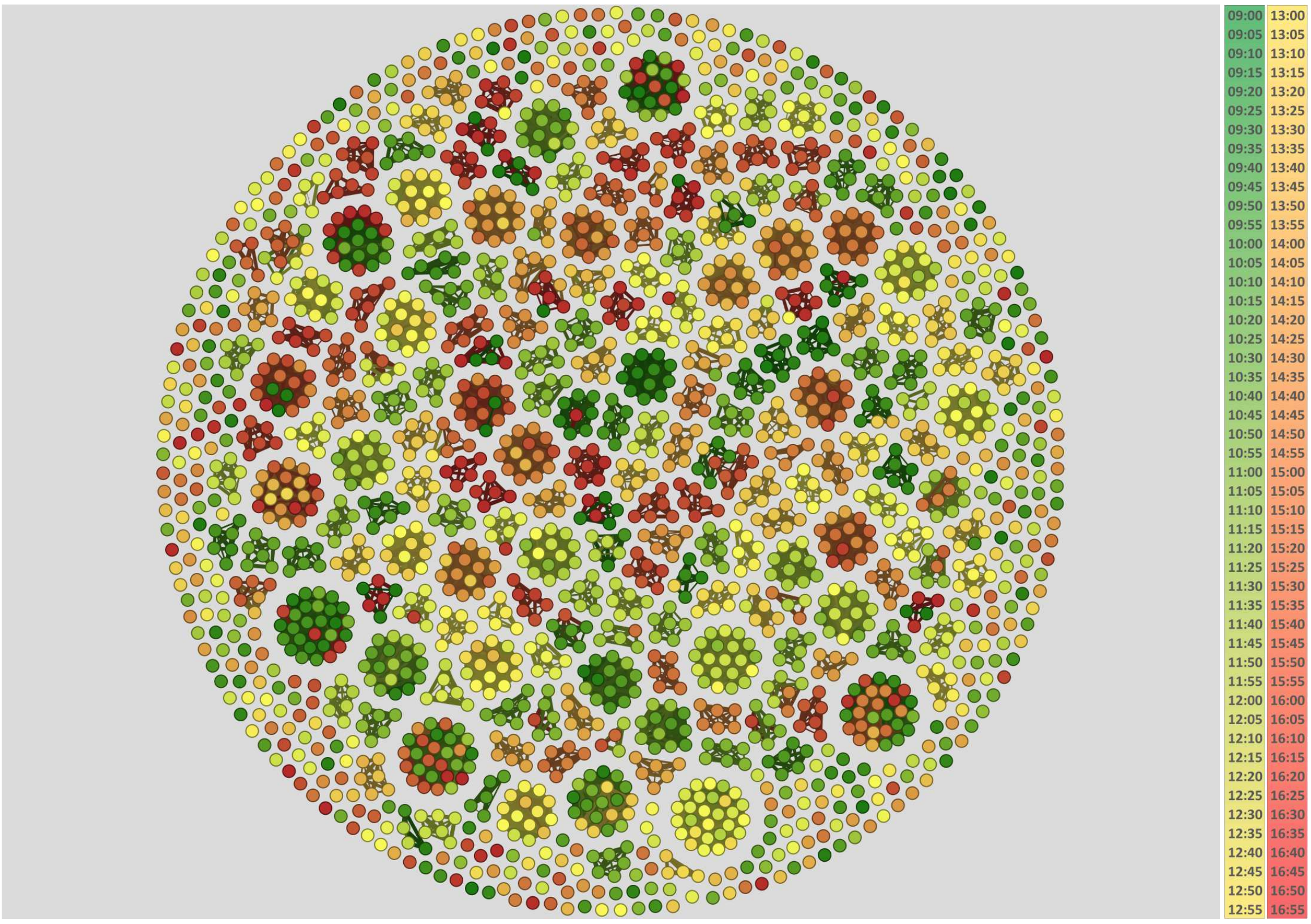

**Figure 9.:** JSE TOP40 5-minute temporal clusters for the period 01-Nov-2012 to 30-Nov-2012, representing 2208 distinct periods. Each node represents a 5-minute period during a trading day, with the colour shading indicating the time-of-day (Morning = green, Lunch = yellow, Afternoon = red) and node connectedness indicating cluster membership.

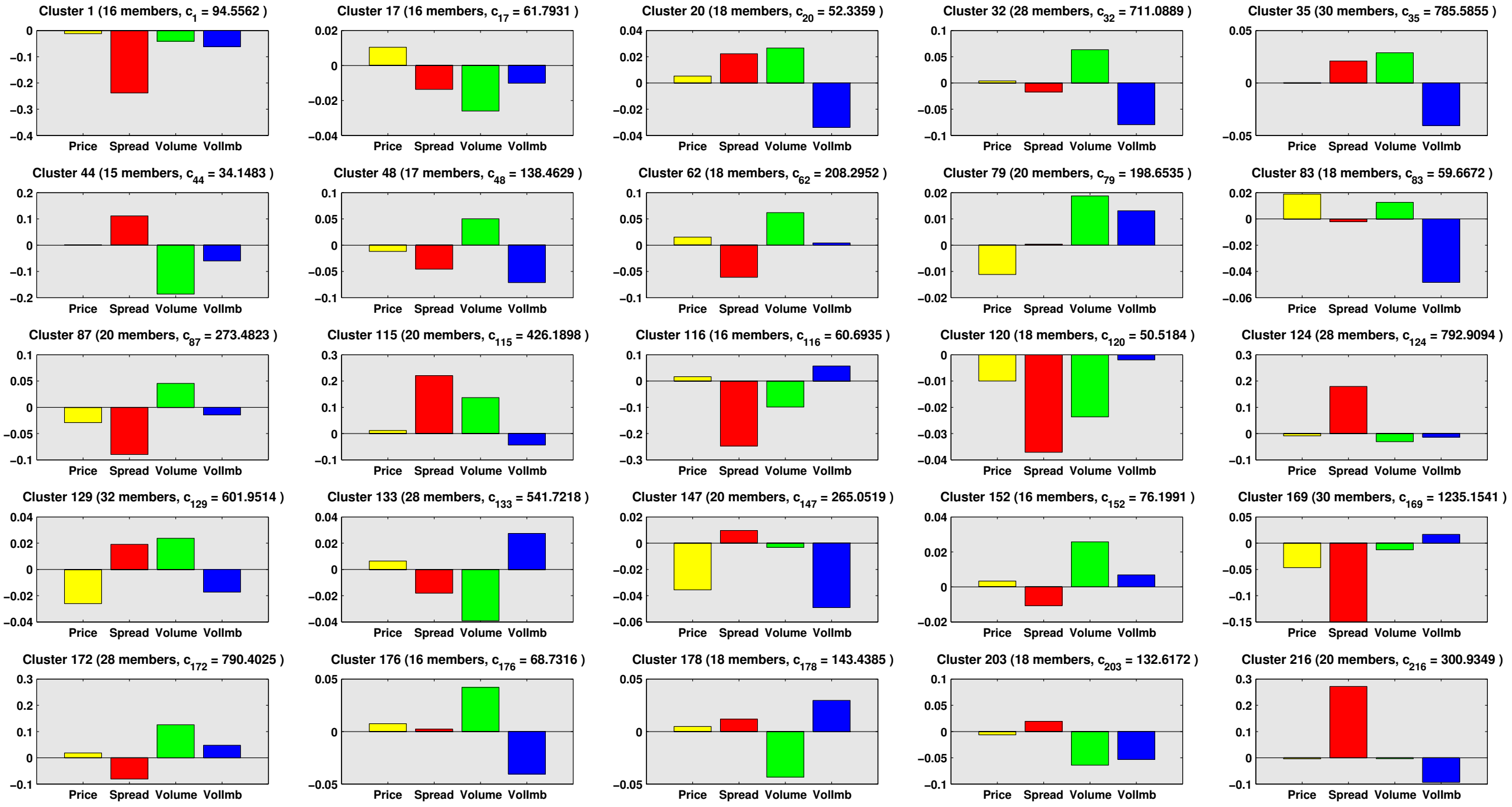

**Figure 10.:** JSE TOP40 5-minute cluster state signature vectors for the period 01-Nov-2012 to 30-Nov-2012. Each plot illustrates the average change in trade price, spread, trade volume and quote volume imbalance across member periods and stocks for each of the clusters with a size $\geq x_{min}$ from the truncated power-law fit. Cluster size and intra-cluster correlation are shown in parentheses.

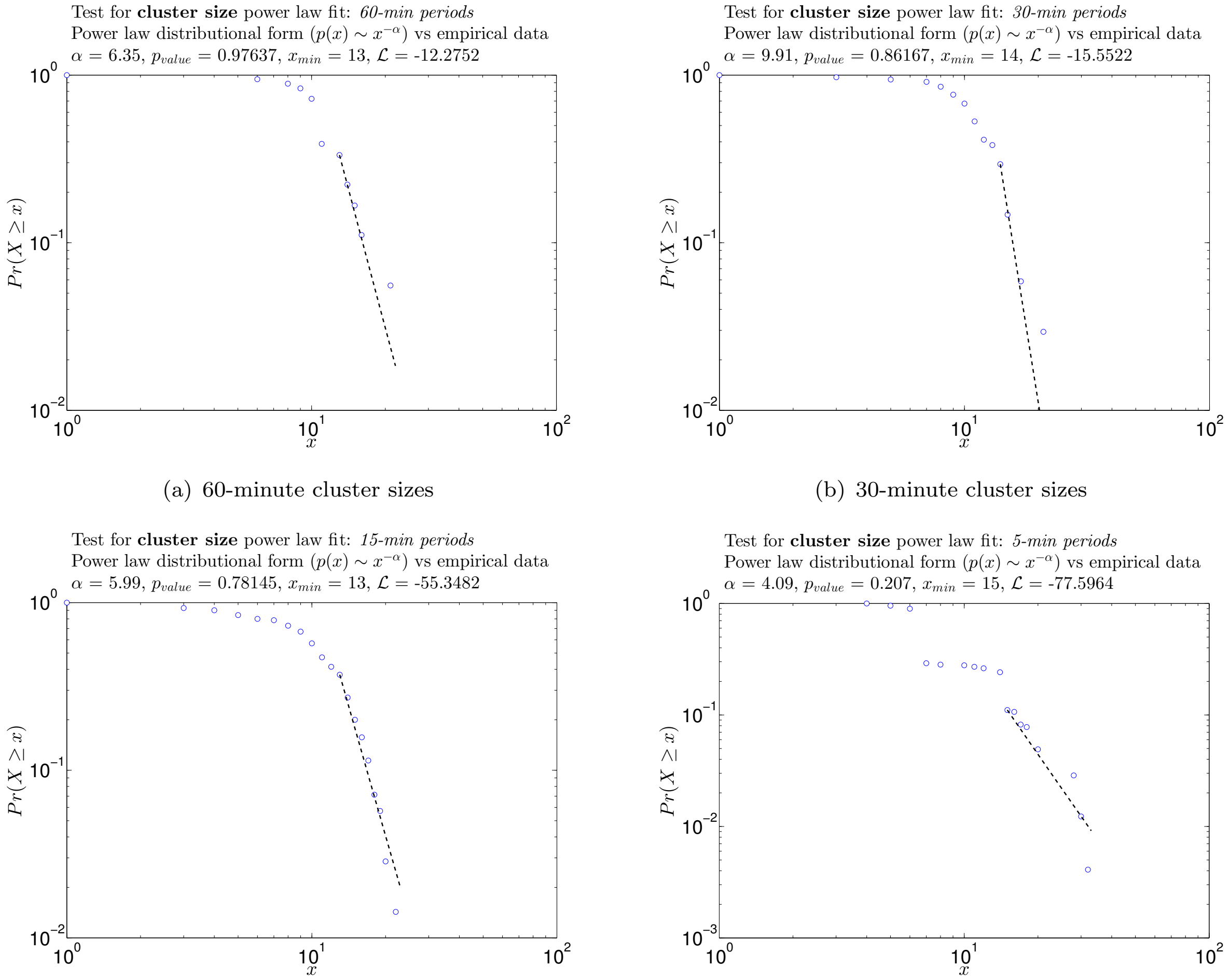

(a) 60-minute cluster sizes

(b) 30-minute cluster sizes

(c) 15-minute cluster sizes

(d) 5-minute cluster sizes

**Figure 11.:** Testing conjecture of power law fit for varying time scale cluster sizes, applying the Clauset, Shalizi and Newman algorithm Clauset et al. (2009). $\alpha$ indicates the scaling parameter of the proposed fit, $p_{value}$ indicates the $p$-value from a Kolmogorov-Smirnov test for the goodness-of-fit of the proposed model to the data, $x_{min}$ indicates the lower-bound for the power law fit and $\mathcal{L}$ is the log-likelihood of the data ($x \geq x_{min}$) under the power law fit.

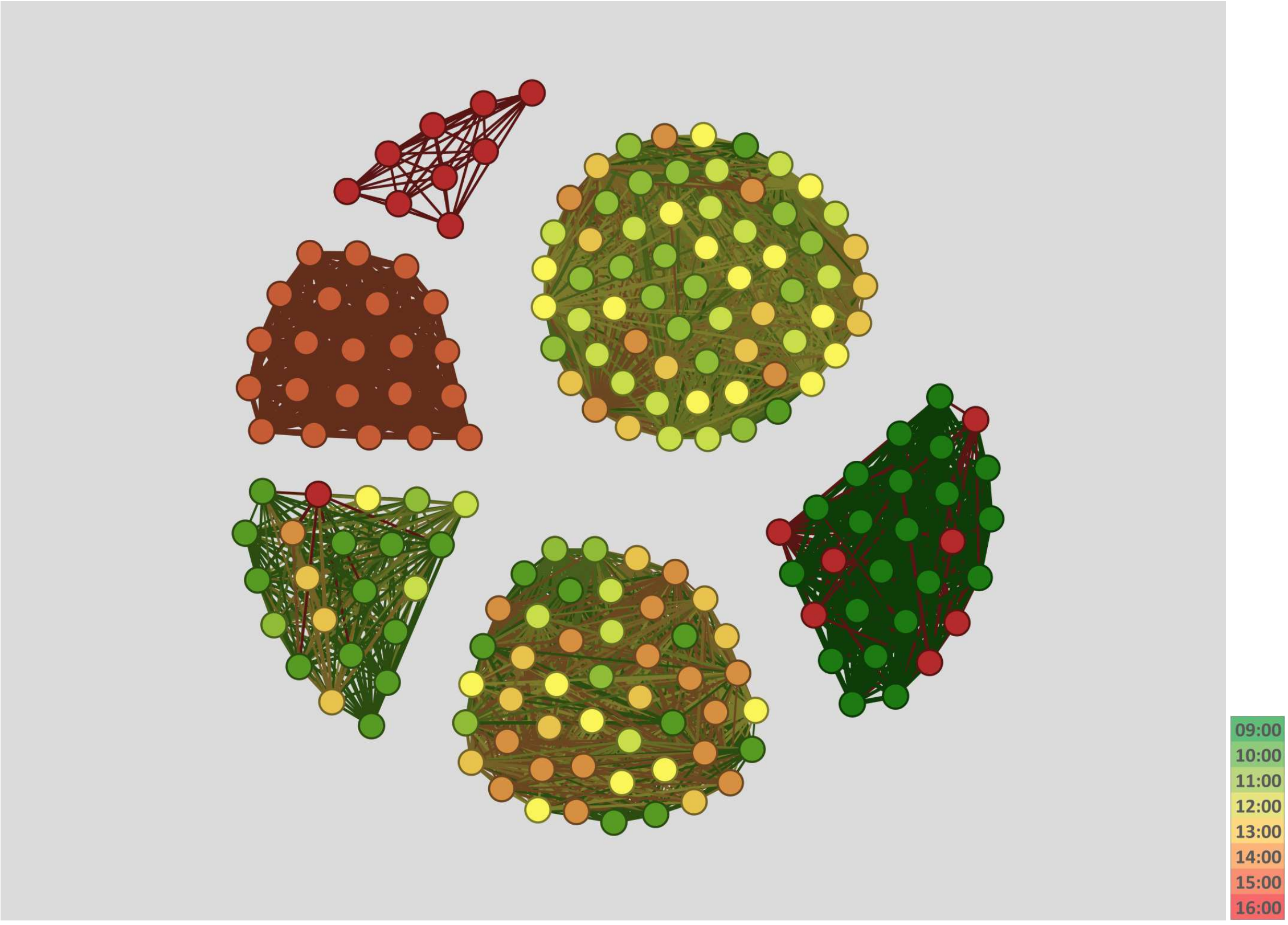

**Figure 12.:** Estimated 60-minute clusters using identified state signature vectors. The Euclidean distance is calculated between each temporal period's feature vector and the state signature vectors. Cluster index assignment is based on the state signature vector which yields the minimum distance.

| | | $\mathbf{state_{t+1}}$ | | | | | |
|---|---|---|---|---|---|---|---|
| | | **1** | **2** | **3** | **4** | **5** | **6** |
| $\mathbf{state_t}$ | **1** | 0.13 | 0.49 | 0.32 | 0.00 | 0.06 | 0.00 |
| | **2** | 0.41 | 0.41 | 0.09 | 0.00 | 0.09 | 0.00 |
| | **3** | 0.00 | 0.00 | 0.00 | 0.52 | 0.05 | 0.43 |
| | **4** | 0.25 | 0.07 | 0.00 | 0.25 | 0.43 | 0.00 |
| | **5** | 0.32 | 0.59 | 0.05 | 0.05 | 0.00 | 0.00 |
| | **6** | 0.00 | 0.00 | 0.00 | 1.00 | 0.00 | 0.00 |

**Table 5.:** Empirical 1-step transition probability matrix for 60-minute states, based on identified temporal cluster configuration. State transitions with a probability $> 0$ are highlighted in green.

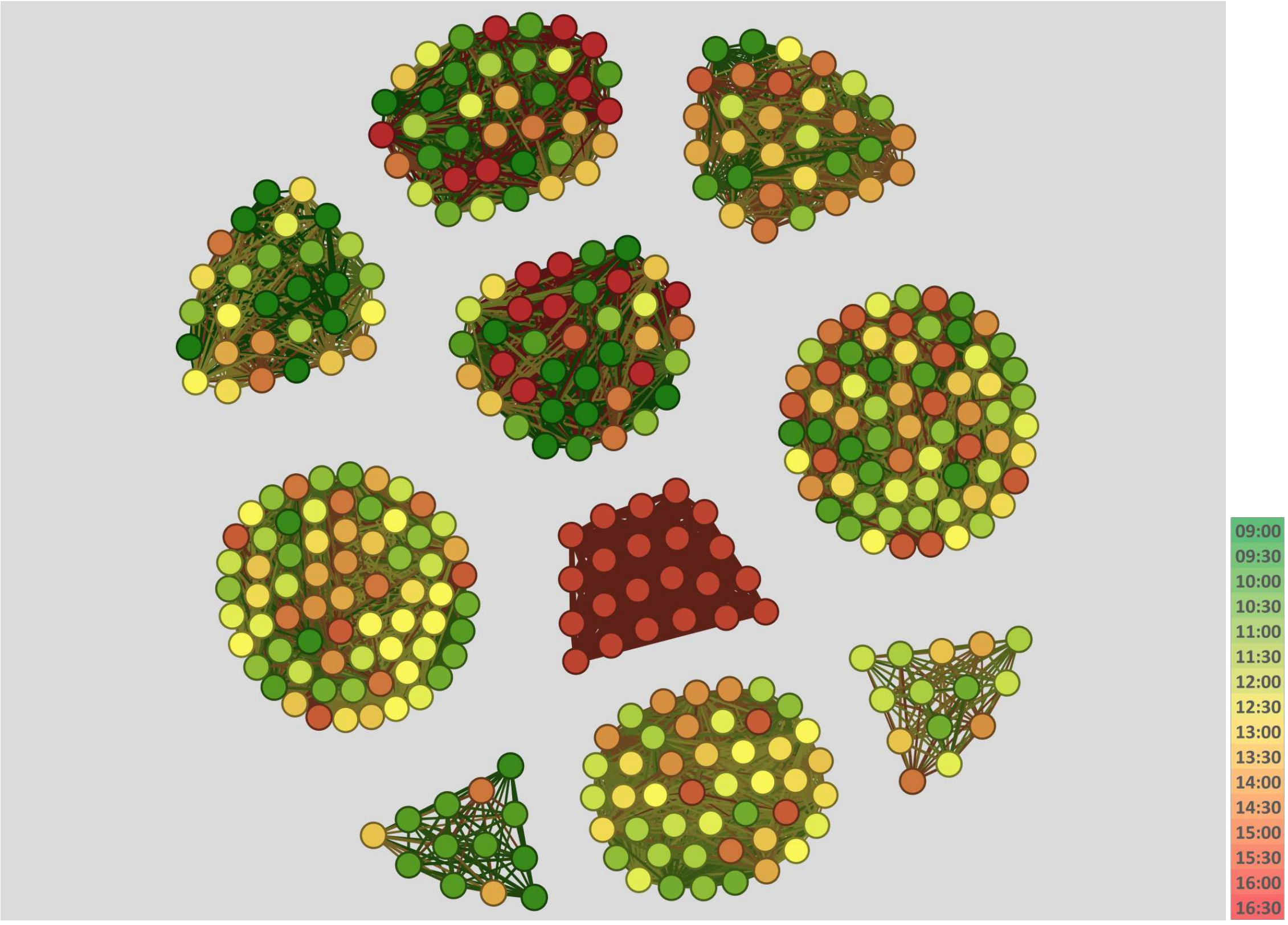

**Figure 13.:** Estimated 30-minute clusters using identified state signature vectors. The Euclidean distance is calculated between each temporal period's feature vector and the state signature vectors. Cluster index assignment is based on the state signature vector which yields the minimum distance.

| | | $\text{state}_{t+1}$ | | | | | | | | | |
|---|---|---|---|---|---|---|---|---|---|---|---|
| | | **1** | **2** | **3** | **4** | **5** | **6** | **7** | **8** | **9** | **10** |
| $\text{state}_t$ | **1** | 0.11 | 0.37 | 0.03 | 0.16 | 0.18 | 0.05 | 0.03 | 0.03 | 0.03 | 0.03 |
| | **2** | 0.07 | 0.04 | 0.35 | 0.06 | 0.04 | 0.10 | 0.17 | 0.10 | 0.04 | 0.01 |
| | **3** | 0.06 | 0.33 | 0.10 | 0.07 | 0.17 | 0.09 | 0.06 | 0.04 | 0.03 | 0.04 |
| | **4** | 0.05 | 0.08 | 0.26 | 0.03 | 0.21 | 0.18 | 0.00 | 0.13 | 0.03 | 0.03 |
| | **5** | 0.07 | 0.13 | 0.24 | 0.11 | 0.04 | 0.09 | 0.07 | 0.13 | 0.04 | 0.07 |
| | **6** | 0.10 | 0.21 | 0.10 | 0.03 | 0.14 | 0.03 | 0.00 | 0.17 | 0.10 | 0.10 |
| | **7** | 0.57 | 0.00 | 0.00 | 0.38 | 0.00 | 0.05 | 0.00 | 0.00 | 0.00 | 0.00 |
| | **8** | 0.13 | 0.23 | 0.16 | 0.16 | 0.13 | 0.03 | 0.06 | 0.06 | 0.03 | 0.00 |
| | **9** | 0.00 | 0.15 | 0.38 | 0.15 | 0.08 | 0.00 | 0.00 | 0.08 | 0.00 | 0.15 |
| | **10** | 0.00 | 0.36 | 0.21 | 0.07 | 0.29 | 0.00 | 0.00 | 0.07 | 0.00 | 0.00 |

**Table 6.:** Empirical 1-step transition probability matrix for 30-minute states, based on identified temporal cluster configuration. State transitions with a probability > 0 are highlighted in green.

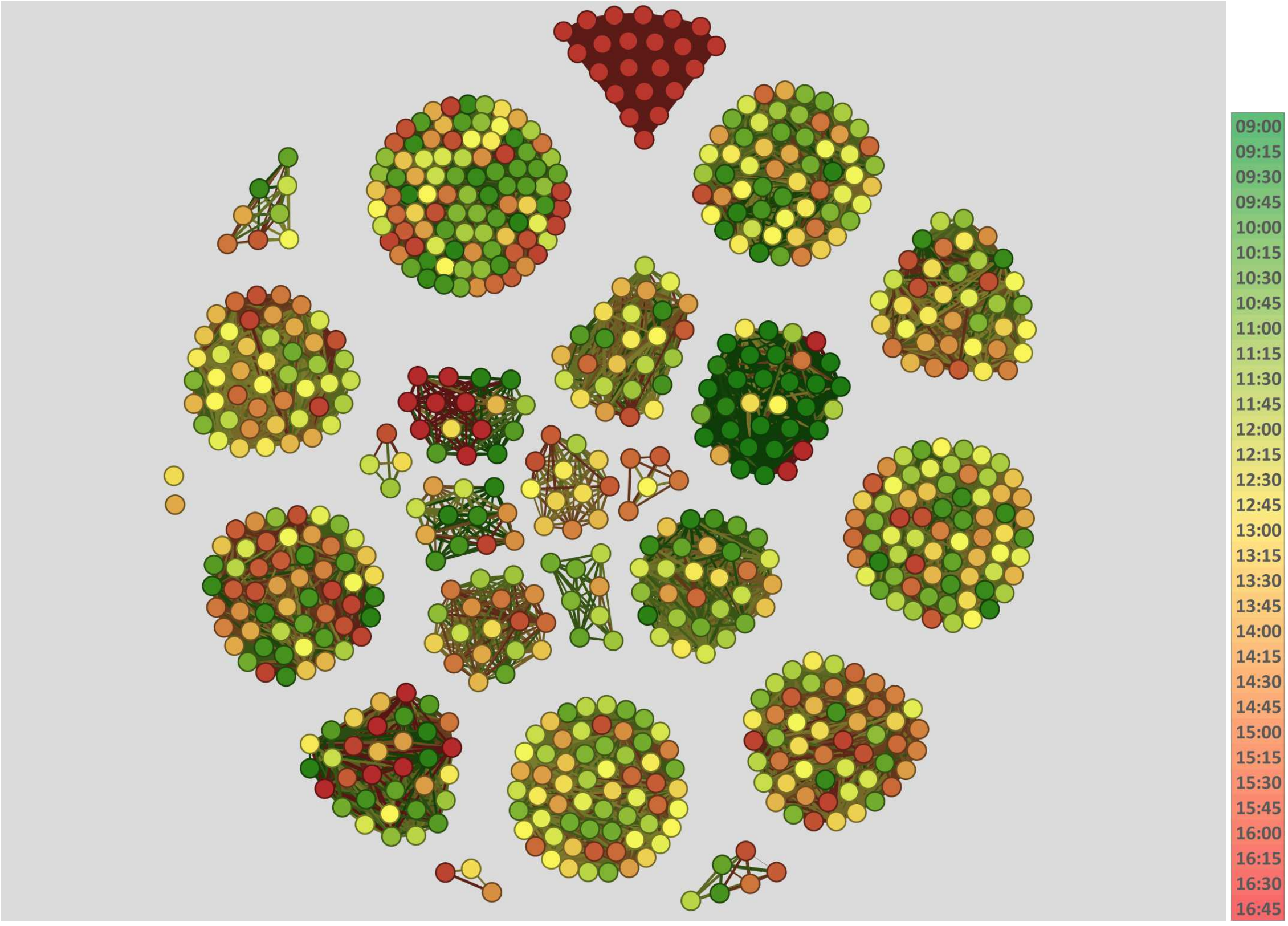

**Figure 14.:** Estimated 15-minute clusters using identified state signature vectors. The Euclidean distance is calculated between each temporal period's feature vector and the state signature vectors. Cluster index assignment is based on the state signature vector which yields the minimum distance.

| | | state$_{t+1}$ | | | | | | | | | | | | | | | | | | | | | | | |
|---|---|---|---|---|---|---|---|---|---|---|---|---|---|---|---|---|---|---|---|---|---|---|---|---|---|
| | | **1** | **2** | **3** | **4** | **5** | **6** | **7** | **8** | **9** | **10** | **11** | **12** | **13** | **14** | **15** | **16** | **17** | **18** | **19** | **20** | **21** | **22** | **23** | **24** | **25** |
| | **1** | 0.00 | 0.20 | 0.13 | 0.00 | 0.07 | 0.00 | 0.03 | 0.17 | 0.10 | 0.07 | 0.03 | 0.03 | 0.03 | 0.03 | 0.00 | 0.00 | 0.00 | 0.03 | 0.07 | 0.00 | 0.00 | 0.00 | 0.00 | 0.00 | 0.00 |
| | **2** | 0.07 | 0.03 | 0.07 | 0.08 | 0.09 | 0.01 | 0.01 | 0.08 | 0.03 | 0.07 | 0.02 | 0.04 | 0.02 | 0.03 | 0.18 | 0.08 | 0.02 | 0.00 | 0.01 | 0.01 | 0.02 | 0.00 | 0.00 | 0.00 | 0.00 |
| | **3** | 0.02 | 0.12 | 0.02 | 0.20 | 0.13 | 0.00 | 0.00 | 0.03 | 0.03 | 0.17 | 0.03 | 0.05 | 0.00 | 0.00 | 0.17 | 0.00 | 0.00 | 0.00 | 0.02 | 0.00 | 0.02 | 0.00 | 0.00 | 0.00 | 0.00 |
| | **4** | 0.07 | 0.16 | 0.02 | 0.07 | 0.09 | 0.00 | 0.02 | 0.13 | 0.00 | 0.04 | 0.00 | 0.09 | 0.05 | 0.00 | 0.02 | 0.07 | 0.00 | 0.00 | 0.07 | 0.05 | 0.00 | 0.02 | 0.02 | 0.00 | 0.00 |
| | **5** | 0.03 | 0.28 | 0.15 | 0.08 | 0.00 | 0.03 | 0.00 | 0.08 | 0.03 | 0.03 | 0.00 | 0.08 | 0.00 | 0.03 | 0.10 | 0.00 | 0.00 | 0.00 | 0.10 | 0.00 | 0.00 | 0.00 | 0.00 | 0.00 | 0.00 |
| | **6** | 0.50 | 0.25 | 0.00 | 0.00 | 0.00 | 0.00 | 0.00 | 0.00 | 0.00 | 0.25 | 0.00 | 0.00 | 0.00 | 0.00 | 0.00 | 0.00 | 0.00 | 0.00 | 0.00 | 0.00 | 0.00 | 0.00 | 0.00 | 0.00 | 0.00 |
| | **7** | 0.00 | 0.17 | 0.17 | 0.00 | 0.00 | 0.00 | 0.17 | 0.33 | 0.17 | 0.00 | 0.00 | 0.00 | 0.00 | 0.00 | 0.00 | 0.00 | 0.00 | 0.00 | 0.00 | 0.00 | 0.00 | 0.00 | 0.00 | 0.00 | 0.00 |
| | **8** | 0.05 | 0.25 | 0.05 | 0.08 | 0.07 | 0.00 | 0.00 | 0.07 | 0.07 | 0.05 | 0.00 | 0.02 | 0.03 | 0.00 | 0.15 | 0.00 | 0.00 | 0.02 | 0.05 | 0.00 | 0.02 | 0.02 | 0.00 | 0.02 | 0.02 |
| | **9** | 0.00 | 0.12 | 0.06 | 0.03 | 0.00 | 0.00 | 0.00 | 0.12 | 0.00 | 0.09 | 0.03 | 0.24 | 0.00 | 0.00 | 0.06 | 0.03 | 0.18 | 0.00 | 0.03 | 0.00 | 0.00 | 0.00 | 0.00 | 0.00 | 0.00 |
| | **10** | 0.04 | 0.10 | 0.19 | 0.00 | 0.02 | 0.00 | 0.00 | 0.19 | 0.02 | 0.04 | 0.00 | 0.02 | 0.04 | 0.02 | 0.12 | 0.06 | 0.06 | 0.00 | 0.02 | 0.02 | 0.02 | 0.02 | 0.00 | 0.00 | 0.00 |
| | **11** | 0.17 | 0.08 | 0.08 | 0.08 | 0.00 | 0.00 | 0.00 | 0.00 | 0.08 | 0.17 | 0.17 | 0.00 | 0.00 | 0.00 | 0.00 | 0.00 | 0.00 | 0.08 | 0.00 | 0.00 | 0.08 | 0.00 | 0.00 | 0.00 | 0.00 |
| **state$_t$** | **12** | 0.04 | 0.09 | 0.09 | 0.11 | 0.09 | 0.00 | 0.00 | 0.09 | 0.04 | 0.06 | 0.00 | 0.00 | 0.04 | 0.02 | 0.17 | 0.04 | 0.04 | 0.02 | 0.00 | 0.02 | 0.02 | 0.02 | 0.00 | 0.00 | 0.00 |
| | **13** | 0.00 | 0.06 | 0.11 | 0.28 | 0.00 | 0.00 | 0.00 | 0.06 | 0.06 | 0.22 | 0.00 | 0.00 | 0.00 | 0.06 | 0.11 | 0.00 | 0.00 | 0.00 | 0.00 | 0.00 | 0.00 | 0.06 | 0.00 | 0.00 | 0.00 |
| | **14** | 0.00 | 0.42 | 0.08 | 0.00 | 0.00 | 0.00 | 0.08 | 0.08 | 0.00 | 0.08 | 0.00 | 0.17 | 0.00 | 0.00 | 0.00 | 0.08 | 0.00 | 0.00 | 0.00 | 0.00 | 0.00 | 0.00 | 0.00 | 0.00 | 0.00 |
| | **15** | 0.01 | 0.18 | 0.15 | 0.01 | 0.01 | 0.01 | 0.01 | 0.07 | 0.03 | 0.04 | 0.01 | 0.21 | 0.07 | 0.00 | 0.03 | 0.03 | 0.00 | 0.00 | 0.07 | 0.03 | 0.00 | 0.00 | 0.00 | 0.00 | 0.00 |
| | **16** | 0.00 | 0.00 | 0.00 | 0.00 | 0.00 | 0.00 | 0.00 | 0.00 | 0.29 | 0.00 | 0.00 | 0.00 | 0.00 | 0.00 | 0.00 | 0.00 | 0.33 | 0.38 | 0.00 | 0.00 | 0.00 | 0.00 | 0.00 | 0.00 | 0.00 |
| | **17** | 0.03 | 0.12 | 0.12 | 0.06 | 0.03 | 0.00 | 0.00 | 0.00 | 0.09 | 0.03 | 0.06 | 0.06 | 0.00 | 0.06 | 0.09 | 0.00 | 0.15 | 0.09 | 0.03 | 0.00 | 0.00 | 0.00 | 0.00 | 0.00 | 0.00 |
| | **18** | 0.00 | 0.06 | 0.06 | 0.06 | 0.06 | 0.00 | 0.00 | 0.00 | 0.00 | 0.06 | 0.06 | 0.00 | 0.06 | 0.00 | 0.00 | 0.00 | 0.50 | 0.00 | 0.06 | 0.00 | 0.00 | 0.00 | 0.00 | 0.00 | 0.00 |
| | **19** | 0.12 | 0.04 | 0.00 | 0.15 | 0.04 | 0.04 | 0.00 | 0.12 | 0.04 | 0.04 | 0.00 | 0.00 | 0.00 | 0.08 | 0.19 | 0.00 | 0.04 | 0.04 | 0.00 | 0.00 | 0.04 | 0.00 | 0.00 | 0.04 | 0.00 |
| | **20** | 0.00 | 0.25 | 0.13 | 0.25 | 0.00 | 0.00 | 0.00 | 0.00 | 0.00 | 0.00 | 0.00 | 0.13 | 0.00 | 0.00 | 0.13 | 0.00 | 0.00 | 0.00 | 0.13 | 0.00 | 0.00 | 0.00 | 0.00 | 0.00 | 0.00 |
| | **21** | 0.00 | 0.13 | 0.00 | 0.13 | 0.00 | 0.00 | 0.00 | 0.00 | 0.00 | 0.38 | 0.00 | 0.13 | 0.00 | 0.00 | 0.13 | 0.00 | 0.00 | 0.00 | 0.13 | 0.00 | 0.00 | 0.00 | 0.00 | 0.00 | 0.00 |
| | **22** | 0.00 | 0.00 | 0.00 | 0.00 | 0.40 | 0.00 | 0.00 | 0.20 | 0.20 | 0.20 | 0.00 | 0.00 | 0.00 | 0.00 | 0.00 | 0.00 | 0.00 | 0.00 | 0.00 | 0.00 | 0.00 | 0.00 | 0.00 | 0.00 | 0.00 |
| | **23** | 0.00 | 0.00 | 0.00 | 0.00 | 0.00 | 0.00 | 0.00 | 0.00 | 1.00 | 0.00 | 0.00 | 0.00 | 0.00 | 0.00 | 0.00 | 0.00 | 0.00 | 0.00 | 0.00 | 0.00 | 0.00 | 0.00 | 0.00 | 0.00 | 0.00 |
| | **24** | 0.00 | 0.00 | 0.00 | 0.00 | 0.00 | 0.00 | 0.00 | 0.33 | 0.00 | 0.33 | 0.00 | 0.00 | 0.00 | 0.00 | 0.00 | 0.33 | 0.00 | 0.00 | 0.00 | 0.00 | 0.00 | 0.00 | 0.00 | 0.00 | 0.00 |
| | **25** | 0.00 | 1.00 | 0.00 | 0.00 | 0.00 | 0.00 | 0.00 | 0.00 | 0.00 | 0.00 | 0.00 | 0.00 | 0.00 | 0.00 | 0.00 | 0.00 | 0.00 | 0.00 | 0.00 | 0.00 | 0.00 | 0.00 | 0.00 | 0.00 | 0.00 |

**Table 7.:** Empirical 1-step transition probability matrix for 15-minute states, based on identified temporal cluster configuration. State transitions with a probability $> 0$ are highlighted in green.

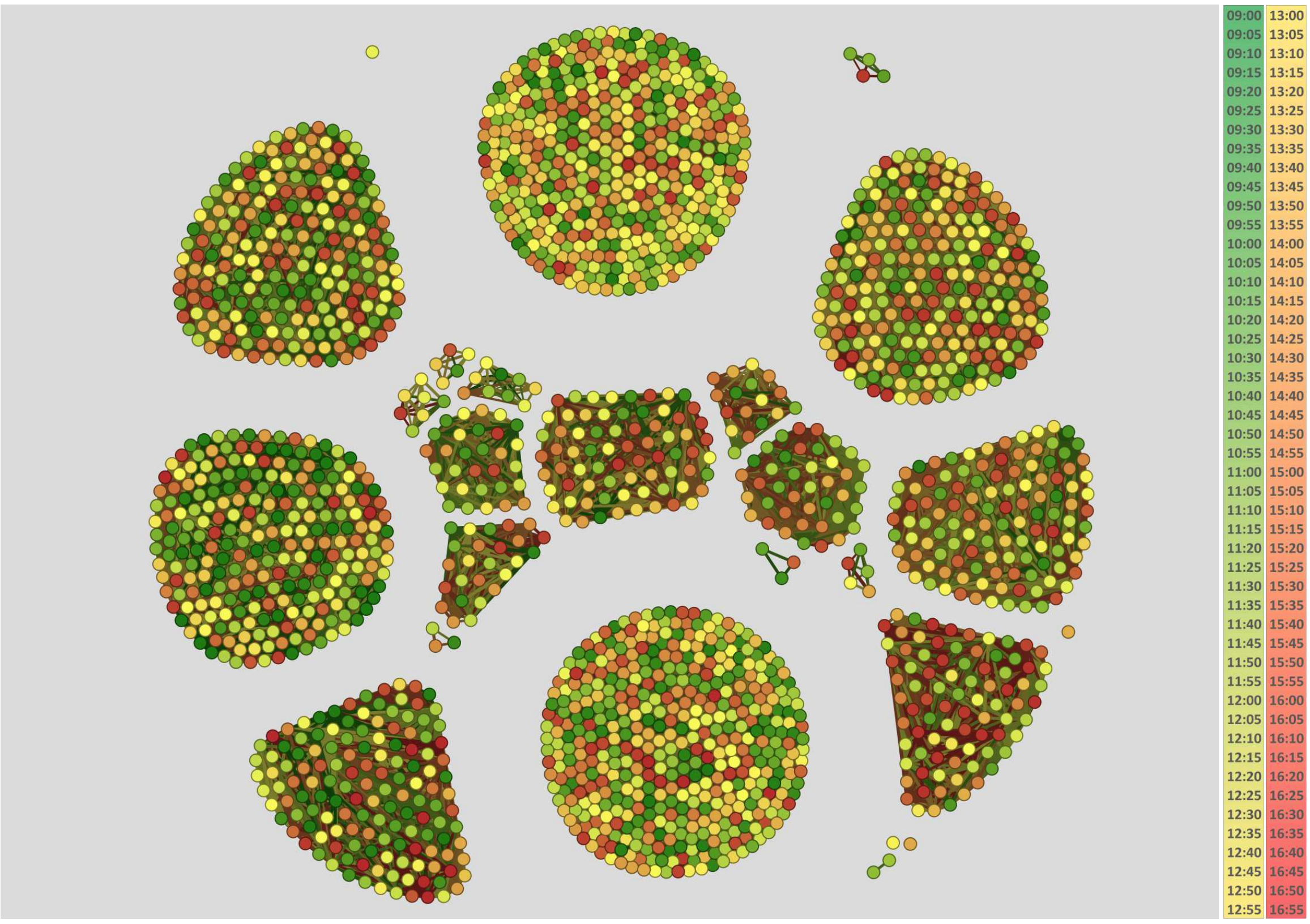

**Figure 15.:** Estimated 5-minute clusters using identified state signature vectors. The Euclidean distance is calculated between each temporal period's feature vector and the state signature vectors. Cluster index assignment is based on the state signature vector which yields the minimum distance.

| | $\mathbf{state_{t+1}}$ | | | | | | | | | | | | | | | | | | | | | | | | |
|---|---|---|---|---|---|---|---|---|---|---|---|---|---|---|---|---|---|---|---|---|---|---|---|---|---|
| | **1** | **2** | **3** | **4** | **5** | **6** | **7** | **8** | **9** | **10** | **11** | **12** | **13** | **14** | **15** | **16** | **17** | **18** | **19** | **20** | **21** | **22** | **23** | **24** | **25** |
| **1** | 0.04 | 0.08 | 0.17 | 0.06 | 0.12 | 0.14 | 0.07 | 0.00 | 0.19 | 0.00 | 0.10 | 0.01 | 0.00 | 0.01 | 0.00 | 0.00 | 0.00 | 0.02 | 0.00 | 0.00 | 0.00 | 0.00 | 0.00 | 0.00 | 0.00 |
| **2** | 0.08 | 0.01 | 0.16 | 0.14 | 0.08 | 0.23 | 0.03 | 0.01 | 0.13 | 0.01 | 0.04 | 0.02 | 0.01 | 0.03 | 0.00 | 0.00 | 0.00 | 0.02 | 0.00 | 0.01 | 0.00 | 0.00 | 0.00 | 0.00 | 0.01 |
| **3** | 0.04 | 0.06 | 0.12 | 0.12 | 0.11 | 0.33 | 0.05 | 0.02 | 0.07 | 0.00 | 0.02 | 0.00 | 0.00 | 0.01 | 0.00 | 0.00 | 0.00 | 0.00 | 0.00 | 0.00 | 0.00 | 0.00 | 0.00 | 0.00 | 0.00 |
| **4** | 0.04 | 0.07 | 0.24 | 0.03 | 0.17 | 0.21 | 0.02 | 0.02 | 0.08 | 0.01 | 0.05 | 0.00 | 0.00 | 0.01 | 0.00 | 0.00 | 0.00 | 0.02 | 0.00 | 0.01 | 0.00 | 0.00 | 0.00 | 0.00 | 0.00 |
| **5** | 0.04 | 0.03 | 0.16 | 0.13 | 0.06 | 0.16 | 0.08 | 0.03 | 0.23 | 0.00 | 0.03 | 0.00 | 0.00 | 0.00 | 0.00 | 0.00 | 0.00 | 0.01 | 0.00 | 0.00 | 0.00 | 0.00 | 0.00 | 0.00 | 0.00 |
| **6** | 0.07 | 0.07 | 0.32 | 0.12 | 0.08 | 0.09 | 0.02 | 0.01 | 0.13 | 0.02 | 0.03 | 0.01 | 0.00 | 0.02 | 0.00 | 0.00 | 0.01 | 0.00 | 0.00 | 0.00 | 0.00 | 0.00 | 0.00 | 0.00 | 0.00 |
| **7** | 0.14 | 0.07 | 0.24 | 0.04 | 0.23 | 0.10 | 0.04 | 0.01 | 0.05 | 0.00 | 0.02 | 0.00 | 0.00 | 0.01 | 0.00 | 0.02 | 0.00 | 0.02 | 0.00 | 0.00 | 0.00 | 0.00 | 0.00 | 0.00 | 0.00 |
| **8** | 0.02 | 0.05 | 0.33 | 0.10 | 0.10 | 0.18 | 0.03 | 0.03 | 0.10 | 0.03 | 0.00 | 0.00 | 0.00 | 0.03 | 0.00 | 0.00 | 0.00 | 0.03 | 0.00 | 0.00 | 0.00 | 0.00 | 0.00 | 0.00 | 0.00 |
| **9** | 0.11 | 0.04 | 0.15 | 0.04 | 0.20 | 0.30 | 0.04 | 0.02 | 0.03 | 0.00 | 0.03 | 0.00 | 0.00 | 0.01 | 0.00 | 0.00 | 0.00 | 0.01 | 0.00 | 0.00 | 0.00 | 0.00 | 0.00 | 0.00 | 0.00 |
| **10** | 0.06 | 0.12 | 0.24 | 0.00 | 0.00 | 0.18 | 0.06 | 0.06 | 0.29 | 0.00 | 0.00 | 0.00 | 0.00 | 0.00 | 0.00 | 0.00 | 0.00 | 0.00 | 0.00 | 0.00 | 0.00 | 0.00 | 0.00 | 0.00 | 0.00 |
| **11** | 0.19 | 0.01 | 0.26 | 0.21 | 0.07 | 0.13 | 0.00 | 0.00 | 0.04 | 0.01 | 0.03 | 0.00 | 0.00 | 0.04 | 0.00 | 0.00 | 0.00 | 0.00 | 0.00 | 0.00 | 0.00 | 0.00 | 0.00 | 0.00 | 0.00 |
| $\mathbf{state_t}$ **12** | 0.11 | 0.11 | 0.00 | 0.00 | 0.22 | 0.22 | 0.00 | 0.22 | 0.11 | 0.00 | 0.00 | 0.00 | 0.00 | 0.00 | 0.00 | 0.00 | 0.00 | 0.00 | 0.00 | 0.00 | 0.00 | 0.00 | 0.00 | 0.00 | 0.00 |
| **13** | 0.00 | 0.33 | 0.33 | 0.00 | 0.00 | 0.00 | 0.00 | 0.00 | 0.00 | 0.00 | 0.00 | 0.00 | 0.00 | 0.00 | 0.00 | 0.00 | 0.00 | 0.33 | 0.00 | 0.00 | 0.00 | 0.00 | 0.00 | 0.00 | 0.00 |
| **14** | 0.03 | 0.03 | 0.17 | 0.00 | 0.10 | 0.37 | 0.00 | 0.03 | 0.10 | 0.00 | 0.07 | 0.00 | 0.00 | 0.07 | 0.00 | 0.00 | 0.00 | 0.00 | 0.00 | 0.00 | 0.00 | 0.00 | 0.03 | 0.00 | 0.00 |
| **15** | 0.00 | 0.00 | 0.33 | 0.00 | 0.33 | 0.00 | 0.00 | 0.33 | 0.00 | 0.00 | 0.00 | 0.00 | 0.00 | 0.00 | 0.00 | 0.00 | 0.00 | 0.00 | 0.00 | 0.00 | 0.00 | 0.00 | 0.00 | 0.00 | 0.00 |
| **16** | 0.00 | 0.00 | 0.20 | 0.40 | 0.20 | 0.00 | 0.00 | 0.20 | 0.00 | 0.00 | 0.00 | 0.00 | 0.00 | 0.00 | 0.00 | 0.00 | 0.00 | 0.00 | 0.00 | 0.00 | 0.00 | 0.00 | 0.00 | 0.00 | 0.00 |
| **17** | 0.20 | 0.00 | 0.20 | 0.20 | 0.00 | 0.40 | 0.00 | 0.00 | 0.00 | 0.00 | 0.00 | 0.00 | 0.00 | 0.00 | 0.00 | 0.00 | 0.00 | 0.00 | 0.00 | 0.00 | 0.00 | 0.00 | 0.00 | 0.00 | 0.00 |
| **18** | 0.00 | 0.12 | 0.24 | 0.24 | 0.04 | 0.12 | 0.04 | 0.08 | 0.04 | 0.00 | 0.00 | 0.00 | 0.00 | 0.00 | 0.00 | 0.00 | 0.00 | 0.04 | 0.00 | 0.00 | 0.00 | 0.04 | 0.00 | 0.00 | 0.00 |
| **19** | 1.00 | 0.00 | 0.00 | 0.00 | 0.00 | 0.00 | 0.00 | 0.00 | 0.00 | 0.00 | 0.00 | 0.00 | 0.00 | 0.00 | 0.00 | 0.00 | 0.00 | 0.00 | 0.00 | 0.00 | 0.00 | 0.00 | 0.00 | 0.00 | 0.00 |
| **20** | 0.14 | 0.14 | 0.14 | 0.14 | 0.00 | 0.14 | 0.00 | 0.00 | 0.29 | 0.00 | 0.00 | 0.00 | 0.00 | 0.00 | 0.00 | 0.00 | 0.00 | 0.00 | 0.00 | 0.00 | 0.00 | 0.00 | 0.00 | 0.00 | 0.00 |
| **21** | 0.00 | 0.00 | 0.00 | 0.00 | 0.00 | 1.00 | 0.00 | 0.00 | 0.00 | 0.00 | 0.00 | 0.00 | 0.00 | 0.00 | 0.00 | 0.00 | 0.00 | 0.00 | 0.00 | 0.00 | 0.00 | 0.00 | 0.00 | 0.00 | 0.00 |
| **22** | 0.00 | 0.00 | 0.25 | 0.00 | 0.00 | 0.50 | 0.00 | 0.00 | 0.25 | 0.00 | 0.00 | 0.00 | 0.00 | 0.00 | 0.00 | 0.00 | 0.00 | 0.00 | 0.00 | 0.00 | 0.00 | 0.00 | 0.00 | 0.00 | 0.00 |
| **23** | 0.00 | 0.00 | 0.50 | 0.50 | 0.00 | 0.00 | 0.00 | 0.00 | 0.00 | 0.00 | 0.00 | 0.00 | 0.00 | 0.00 | 0.00 | 0.00 | 0.00 | 0.00 | 0.00 | 0.00 | 0.00 | 0.00 | 0.00 | 0.00 | 0.00 |
| **24** | 0.00 | 0.00 | 0.00 | 1.00 | 0.00 | 0.00 | 0.00 | 0.00 | 0.00 | 0.00 | 0.00 | 0.00 | 0.00 | 0.00 | 0.00 | 0.00 | 0.00 | 0.00 | 0.00 | 0.00 | 0.00 | 0.00 | 0.00 | 0.00 | 0.00 |
| **25** | 0.00 | 0.00 | 0.00 | 0.00 | 0.00 | 0.00 | 0.00 | 0.00 | 0.00 | 0.00 | 0.00 | 0.00 | 0.00 | 1.00 | 0.00 | 0.00 | 0.00 | 0.00 | 0.00 | 0.00 | 0.00 | 0.00 | 0.00 | 0.00 | 0.00 |

**Table 8.:** Empirical 1-step transition probability matrix for 5-minute states, based on identified temporal cluster configuration. State transitions with a probability $> 0$ are highlighted in green.

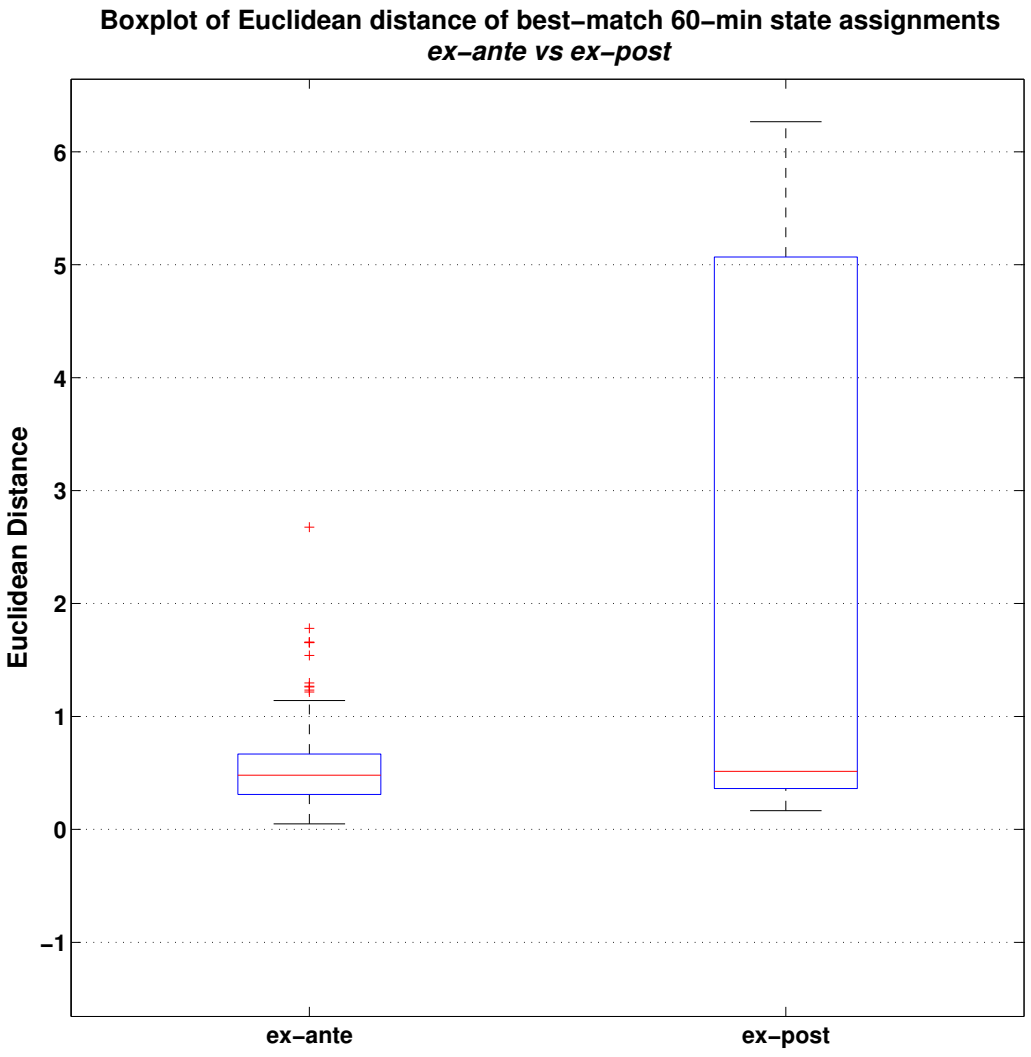

(a) 60-minute states

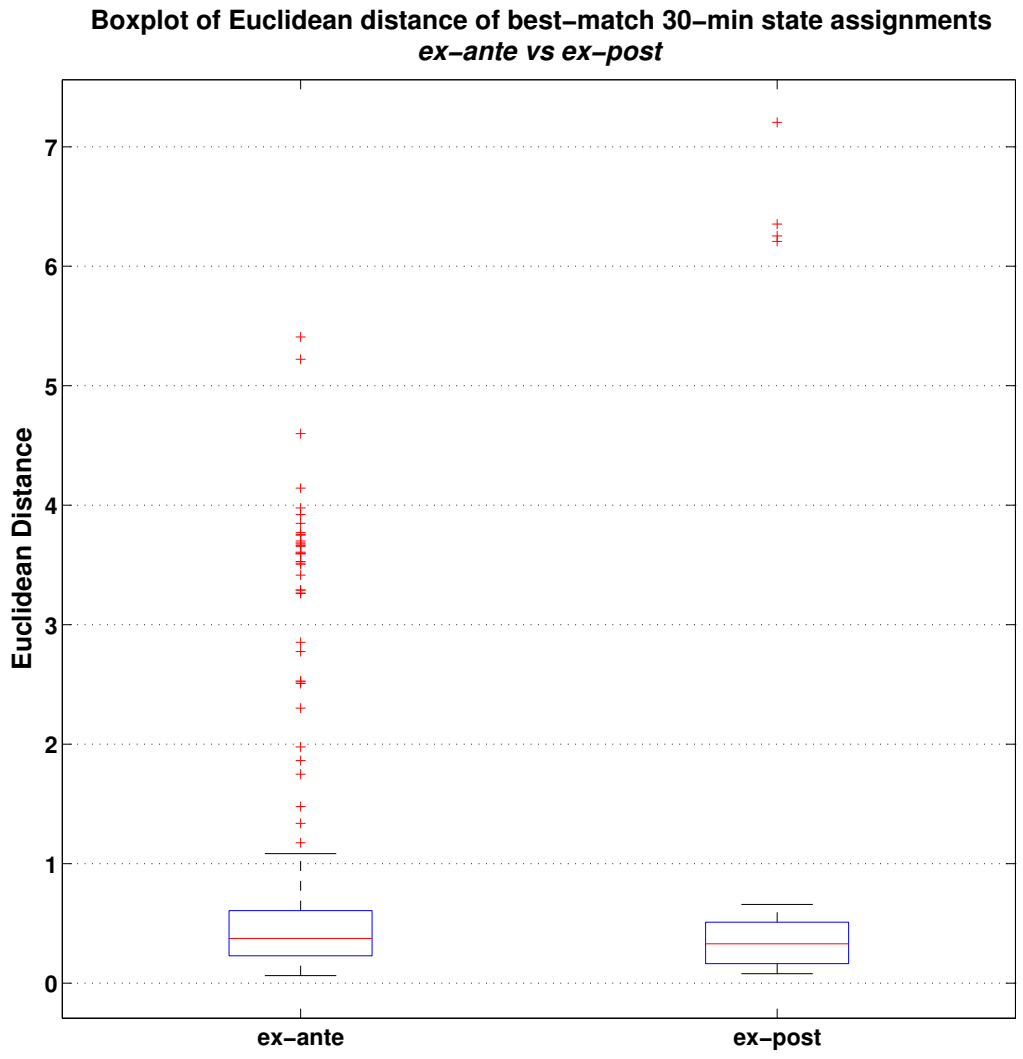

(b) 30-minute states

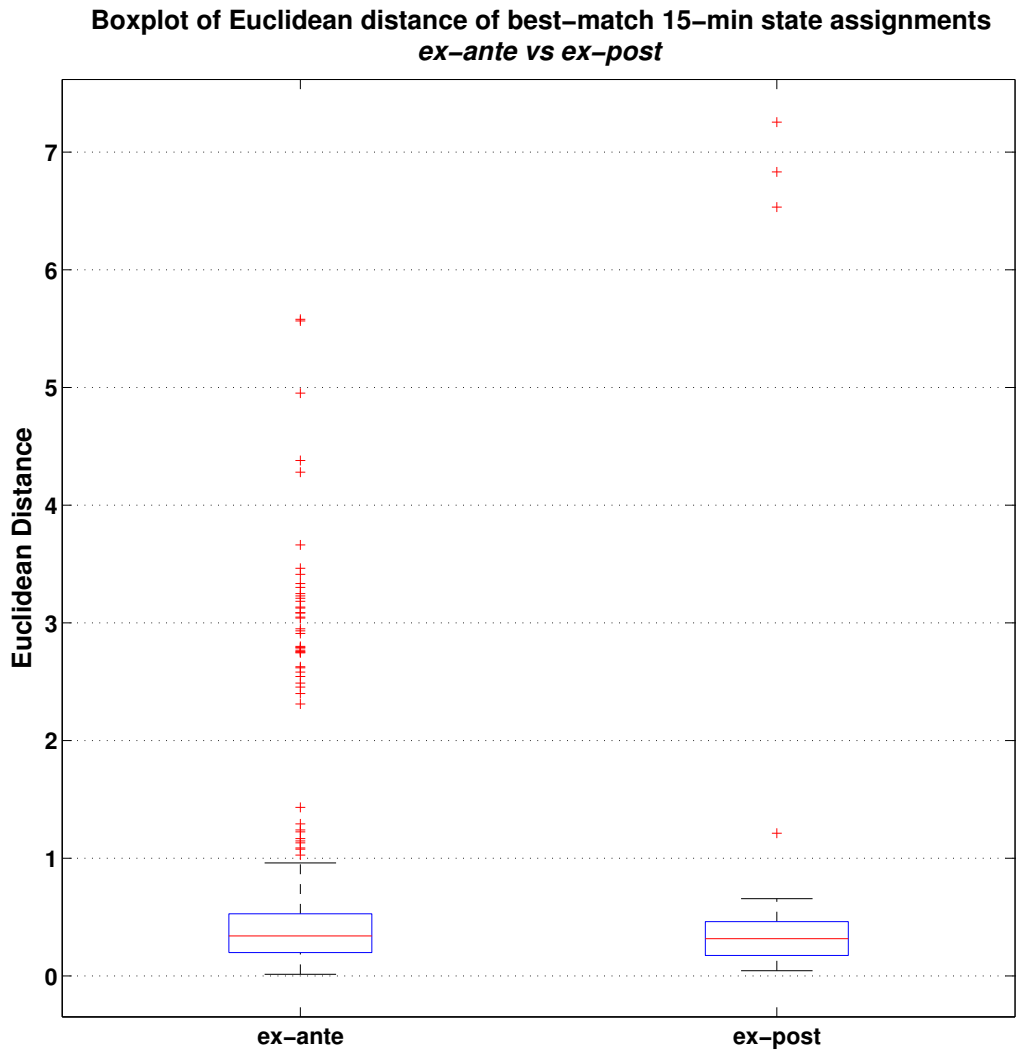

(c) 15-minute states

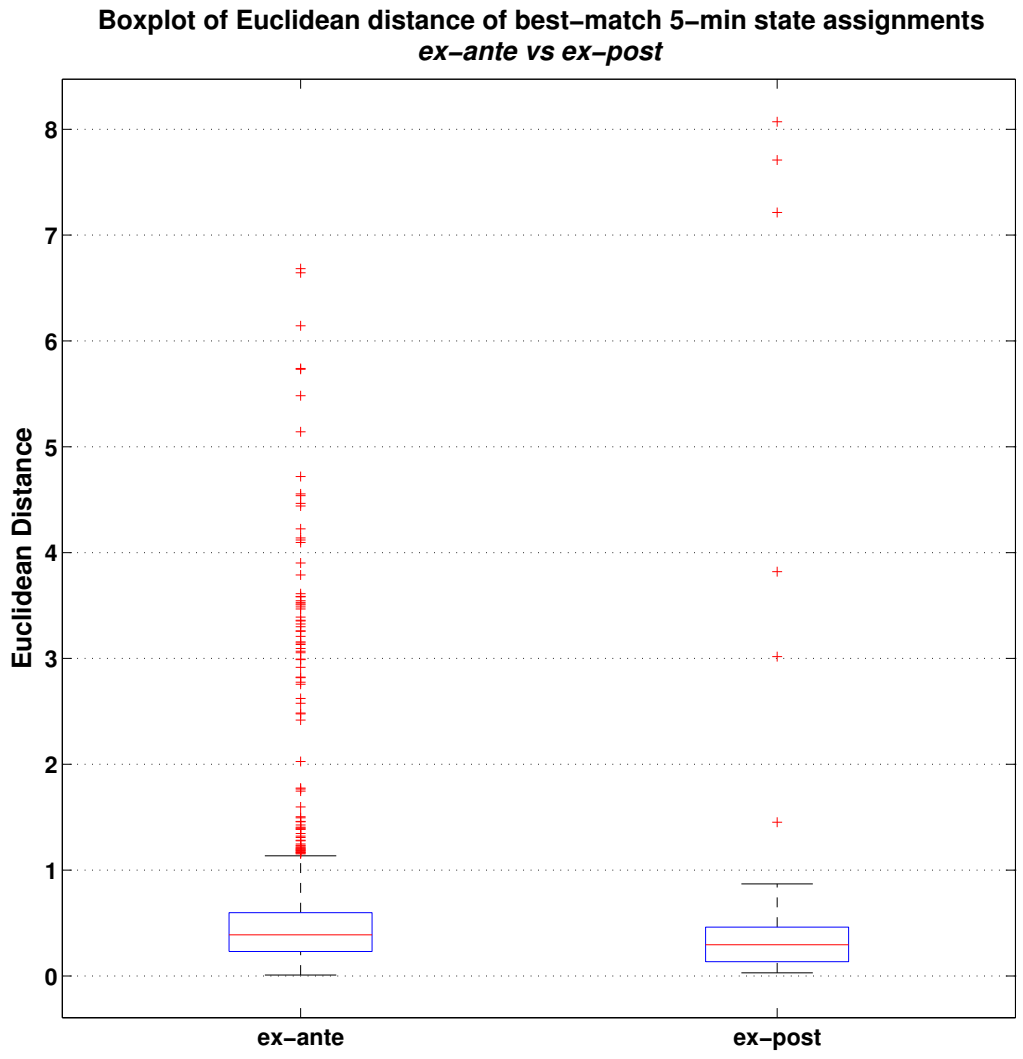

(d) 5-minute states

**Figure 16.:** Measuring the stability of the online state assignment algorithm out-of-sample. Given that the state assignment of an online FV is based on the minimum Euclidean distance to predetermined SSVs, we compute the *best match* distance for each of the FVs in a sample and use a boxplot to visualise the empirical distribution. In this figure, we compare the *ex-ante* (01-Nov-2012 to 30-Nov-2012, same period used for SSV estimation) and *ex-post* (03-Dec-2012 to 07-Dec-2012, one week after SSV estimation window) periods.

## 9. Conclusion

In this paper, we have outlined a novel approach for the unsupervised detection of intraday temporal market states at varying time scales, as well as a proposed mechanism for significant state selection and online state estimation. Using the maximum likelihood approach of Giada and Marsili (2001), we show that the technique can be used to cluster temporal periods as objects based on market microstructure feature performance. A high-speed PGA was used for cluster detection, with a computation time conducive to overnight or even intraday calibration of market states. A study of temporal cluster configurations and power-law fits to 60-minute, 30-minute, 15-minute and 5-minute time scales revealed scale-specific system behaviour, motivating the need for scale-specific state space reduction for optimal planning of participating trading agents. The proposed scheme for online state detection suggested the use of SSVs to capture the market activity signature of each identified state, with a simple distance metric of the prevailing FV to determine the state index. We showed that the online state detection scheme can be used to enumerate and update 1-step transition probability matrices, which can be used for optimal planning in the high-frequency trading domain. We considered the stability of the algorithm *ex-post* and found that we could reliably determine 30-minute, 15-minute and 5-minute states using the proposed algorithm, whereas 60-minute states were less stable.

While this paper demonstrates a feasible framework for temporal state detection, further research should consider a longer-term study to determine the stability of identified states and explore alternative propositions for features, state signature extraction and online detection. In the South African equity market, the impact of significant infrastructure changes (e.g. exchange server migration, fee model modifications, co-located trading servers) on temporal system behaviour can be considered.

## Acknowledgements

This work is based on the research supported in part by the National Research Foundation of South Africa (Grant number 89250). The conclusions herein are due to the authors and the NRF accepts no liability in this regard. We also thank the Fields Institute for Research in Mathematical Sciences and the University of Toronto for hosting the first author while much of the work for this paper was completed.

## References

Abergel, F. and Jedidi, A. (2015). Long time behaviour of a hawkes process-based limit order book. *Working paper, Available at SSRN: http://ssrn.com/abstract=2575498.*

Adi, A., Botzer, D., Nechushtai, G., and Sharon, G. (2006). Complex event processing for financial services. *Proceedings from the IEEE Services Computing Workshops*, pages 7–12.

Arthur, W. (1995). Complexity in economic and financial markets. *Complexity*, **1**(1):20–25.

Arthur, W., Holland, J., LeBaron, B., Palmer, R., and Taylor, P. (1997). Asset pricing under endogenous expectations in an artificial stock market. *The Economy as an Evolving Complex System*, **2**:15–44.

Bacry, E., Mastromatteo, I., and Muzy, J. (2015). Hawkes processes in finance. *Market Microstructure and Liquidity.*

Baldovin, F., Camana, F., Caporin, M., Caraglio, M., and Stella, A. (2015). Ensemble properties of high-frequency data and intraday trading rules. *Quantitative Finance*, **15**(2):231–245.

Bastian, M., Heymann, S., and Jacomy, M. (2009). Gephi: an open source software for exploring and manipulating networks. *Proceedings from the International AAAI Conference on Weblogs and Social Media.*

Bauke, H. (2007). Parameter estimation for power-law distributions by maximum likelihood methods. *The European Physical Journal B*, **58**(2):167–173.

Biais, L., Glosten, C., and Spatt, C. (2005). Market microstructure: A survey of microfoundations, empirical results, and policy implications. *Journal of Financial Markets*, **8**(2):217–264.

Blatt, M., Wiseman, S., and Domany, E. (1996). Superparamagnetic clustering of data. *Phys. Rev. Lett.*, **76**(18):3251–3254.

Blatt, M., Wiseman, S., and Domany, E. (1997). Data clustering using a model granular magnet. *Neural Computation*, **9**:1805–1842.

Brock, W. (1993). Pathways to randomness in the economy: emergent nonlinearity and chaos in economics and finance. *Estudios Economicos*, **8**:3–55.

Cieslakiewicz, D. (2014). Unsupervised asset cluster analysis implemented with parallel genetic algorithms on the nvidia cuda platform. Master's thesis, University of the Witwatersrand.

Clauset, A., Shalizi, C., and Newman, M. (2009). Power-law distributions in empirical data. *SIAM Review*, **51**(4):661–703.

Cont, R. and Tankov, P. (2004). *Financial modelling with jump processes*. Chapman & Hall, CRC Financial Mathematics Series.

Dacorogna, M., Gauvreau, C., Muller, U., Olsen, R., and Pictet, O. (1996). Changing time scale for short-term forecasting in financial markets. *Journal of Forecasting*, **15**:203–227.

Derman, E. (2002). The perception of time, risk and return during periods of speculation. *Quantitative Finance*, **2**:282–296.

Easley, D., López de Prado, M., and O'Hara, M. (2012). The volume clock: Insights into the high-frequency paradigm (digest summary). *Journal of Portfolio Management*, **39**(1):19–29.

Emmert-Streib, F. and Dehmer, M. (2010). Influence of the time scale on the construction of financial networks. *PLoS ONE*, **5**(9).

Engle, R. and Russell, J. (1998). Autoregressive conditional duration: A new model for irregularly spaced transaction data. *Econometrica*, **66**:11271162.

Fruchterman, T. and Reingold, E. (1991). Graph drawing by force-directed placement. *Software - practice and experience*, **21**(11):1129–1164.

Gabaix, X., Gopikrishnan, P., Plerou, V., and Stanley, H. (2003). A theory of power-law distributions in financial market fluctuations. *Nature*, **423**(6937):267–270.

Garman, M. (1976). Market microstructure. *Journal of Financial Economics*, **3**:257–275.

Gençay, R., Gradojevic, N., Selçuk, F., and Whitcher, B. (2010). Asymmetry of information flow between volatilities across time scales. *Quantitative Finance*, **10**(8):895–915.

Giada, L. and Marsili, M. (2001). Data clustering and noise undressing of correlation matrices. *Phys. Rev. E*, **63**(1).

Hasbrouck, J. (1988). Trades, quotes, inventories and information. *Journal of Financial Economics*, **22**:229–252.

Hasbrouck, J. (1991). Measuring the information content of stock trades. *Journal of Finance*, **46**:179–207.

Hendricks, D., Gebbie, T., and Wilcox, D. (2015). High-speed detection of emergent market clustering via an unsupervised parallel genetic algorithm. *South African Journal of Science (to appear)*.

Hinton, G. (2007). Learning multiple layers of representation. *Trends in Cognitive Sciences*, **11**(10):428–434.

Hommes, C. (2001). Financial markets as nonlinear adaptive evolutionary systems. *Quantitative Finance*, **1**(1):149–167.

JSE (2014). Dual-listed companies (retrieved: 08/03/2014). url: http://www.jse.co.za/how-tolist/main-board/dual-listed-companies.aspx.

JSE (2015). Market data - equities, derivatives and interest rate products price list (retrieved: 13/07/2015).

Kullmann, L., Kertész, J., and Mantegnae, R. (2000). Identification of clusters of companies in stock indices via potts super-paramagnetic transitions. *Working paper, Available at arXiv: http://arxiv.org/abs/cond-mat/0002238*.

Large, J. (2000). Measuring the resiliency of an electronic limit order book. *Journal of Financial Markets*, **10**:1–25.

Madhavan, A. (2000). Market microstructure: A survey. *Journal of Financial Markets*, **3**(3):205–258.

Marsili, M. (2002). Dissecting financial markets: sectors and states. *Quantitative Finance*, **2**(4):297–302.

Mastromatteo, I. and Marsili, M. (2011). On the criticality of inferred models. *Journal of Statistical Mechanics: Theory and Experiment*, **2011**(10).

McLachlan, G., Peel, D., and Whiten, W. (1996). Maximum likelihood clustering via normal mixture models. *Signal Processing: Image Communication*, **8**(2):105–111.

McNeil, A., Frey, R., and Embrechts, P. (2015). *Quantitative Risk Management: Concepts, Techniques and*

*Tools*. Princeton University Press, Princeton Series in Finance.

Müller, U., Dacorogna, M., Davé, R., Pictet, O.V. Olsen, R., and Ward, J. (1995). Fractals and intrinsic time a challenge to econometricians. *Olsen and Associates, Zurich*.

Mungan, M. and Ramasco, J. (2010). Stability of maximum-likelihood-based clustering methods: exploring the backbone of classifications. *Journal of Statistical Mechanics: Theory and Experiment*, **4**.

Noh, J. (2000). A model for correlations in stock markets. *Physical Review E*, **61**.

O'Hara, M. (1998). *Market Microstructure Theory*. Blackwell publishing.

Patterson, D. and Hennessy, J. (2013). *Computer Organization and Design: The Hardware/Software Interface, Fifth edition*. Morgan Kaufmann.

Toke, I. and Pomponio, F. (2012). Modelling trades-through in a limit order book using hawkes processes. *Economics: The Open-Access, Open-Assessment E-Journal*, **6**(22):1–23.

Wang, S. and Swendsen, R. (1990). Cluster monte carlo algorithms. *Physica A*, **167**(565).

Wilcox, D. and Gebbie, T. (2014). Hierarchical causality in financial economics. *Working paper, Available at SSRN: http://ssrn.com/abstract=2544327*.

Wiseman, S., Blatt, M., and Domany, E. (1998). Superparamagnetic clustering of data. *Phys. Rev. E*, **57**:37–67.

Zhang, L., Mykland, P., and Aït-Sahalia, Y. (2005). A tale of two time scales: Determining integrated volatility with noisy high-frequency data. *Journal of the American Statistical Association*, **100**(472):1394–1411.

## Appendix A: The Noh-Giada-Marsili coupling parameters

According to Noh (2000), the generative model of the price associated with the $i^{th}$ stock can be written as

$$X_i(t) = g_{s_i}\eta_{s_i} + \sqrt{1 - g_{s_i}^2}\epsilon_i, \tag{A1}$$

where the cluster-related influences are driven by $\eta_{s_i}$ and the stock-specific influences by $\epsilon_i$. Both innovations are treated as Gaussian random variables with unit variance and zero mean[1]. The relative contribution is controlled by the intra-cluster coupling parameter $g_{s_i}$. The Noh-Giada-Marsili model encodes the idea that stocks which have something in common belong in the same cluster. This comes with the caveat that stock membership in clusters is mutually exclusive and intra-cluster correlations are positive.

From Equation A1 we compute the covariance for the $i^{th}$ and $j^{th}$ stocks

$$\mathrm{E}[X_i(t)X_j(t)] = g_{s_i}^2\mathrm{E}[\eta_{s_i}\eta_{s_j}] + (1 - g_{s_i}^2)\mathrm{E}[\epsilon_i\epsilon_j]. \tag{A4}$$

Using the assumption of unit variance and zero mean for both the shared component ($\eta_{s_i}$) and stock component ($\epsilon_i$) processes, the correlation between stock $i$ and $j$ is given by

$$C_{ij} = g_{s_i}^2\delta_{s_i s_j} + (1 - g_{s_i}^2)\delta_{ij}. \tag{A5}$$

The following cluster relations can be derived, where $n_s$ is the index of stock in the $s^{th}$ cluster and $c_s$ is the internal correlation of the $s^{th}$ cluster, given that clusters are mutually exclusive

$$n_s = \sum_{i=1}^{N}\delta_{s_i s}, \qquad c_s = \sum_{i,j=1}^{N} C_{ij}\delta_{s_i s}\delta_{s_j s}. \tag{A6}$$

From Equation A5, for $s_i = s_j = s$, we have $C_{ij} \approx g_s^2$ (Giada and Marsili (2001)). We can multiply both sides of Equation A5 by $\delta_{s_i s}\delta_{s_j s}$ and sum over all $i$ and $j$ to find

$$\sum_{i,j} C_{ij}\delta_{s_i s}\delta_{s_j s} = \sum_{i,j} g_{s_i}^2\delta_{s_i s_j}\delta_{s_i s}\delta_{s_j s} + \sum_{i,j}(1 - g_{s_i}^2)\delta_{ij}\delta_{s_i s}\delta_{s_j s}. \tag{A7}$$

To sum out the delta functions over the clusters and stocks from

$$\sum_{i,j} C_{ij}\delta_{s_i s}\delta_{s_j s} = \sum_{i}(g_{s_i}^2\delta_{s_i s}\sum_{j}\delta_{s_i s_j}\delta_{s_j s}) + \sum_{i}((1 - g_{s_i}^2)\delta_{s_i s}\sum_{j}\delta_{ij}\delta_{s_j s}), \tag{A8}$$

[1]This form of the price model ensures that the self correlation of a stock is one and independent of the cluster coupling. This can be seen by computing the self correlation $E[x_i^2]$ and using that clusters and stock unique process are unit variance zero mean processes

$$\mathrm{E}[(g_{s_i}\eta_{s_i} + \sqrt{1 - g_{s_i}^2}\epsilon_i)^2] = g_{s_i}^2 + (1 - g_{s_i}^2) = 1. \tag{A2}$$

This is not a unique choice, another possible choice often used is

$$\mathrm{E}[(\frac{\sqrt{g_{s_i}}}{\sqrt{1 + g_{s_i}}}\eta_{s_i} + \frac{1}{\sqrt{1 + g_{s_i}}}\epsilon_i)^2] = \frac{1 + g_{s_i}}{1 + g_{s_i}} = 1. \tag{A3}$$

we use that $\sum_j \delta_{ij}\delta_{s_i s} = \delta_{s_i s}$, $\sum_j \delta_{s_i s_j}\delta_{s_j s} = n_s \delta_{s_i s}$ and $\sum_i \delta_{s_i s}^2 = \sum_i \delta_{s_i s}$ to find

$$\sum_{i,j} C_{ij}\delta_{s_i s}\delta_{s_j s} = g_s^2 n_s \sum_i \delta_{s_i s} + (1-g_s^2)\sum_i \delta_{s_i s}. \tag{A9}$$

By combining Equations A6 and A9, we get

$$c_s = g_s^2 n_s^2 + (1-g_s^2) n_s = g_s^2(n_s^2 - n_s) - n_s. \tag{A10}$$

This is can be rearranged to finally obtain an expression for the intra-cluster coupling parameter for cluster $s$,

$$g_s = \sqrt{\frac{c_s - n_s}{n_s^2 - n_s}}. \tag{A11}$$

**Appendix B: The Noh-Giada-Marsili likelihood function**

We evaluate the probability of the data satisfying the model by using the multiplicative property of probabilities,

$$P(X_1(1), \ldots, X_N(D)) = \prod_{d=1}^{D}\prod_{i=1}^{N} P(X_i(d)). \tag{B1}$$

The probability of being in a given state that satisfies the model is given as a delta function, such that we sum over all $N$ stocks and all $D$ features (date-times), taking expectations $\langle \ldots \rangle_{\eta,\epsilon}$ over the random processes associated with the stock-specific noise and the cluster-specific noise

$$P = \prod_{d=1}^{D} \left\langle \prod_{i=1}^{N} \delta\left( X_i(d) - (g_{s_i}\eta_{s_i} + \sqrt{1-g_{s_i}^2}\epsilon_i)\right) \right\rangle_{\eta,\epsilon}. \tag{B2}$$

This takes on the form

$$P = \prod_{d=1}^{D}\prod_{i=1}^{N} \int d\epsilon_i d\eta_{s_i} \exp\left[ -\frac{1}{2}\sum_{k}^{N} \epsilon_k \delta_{ki}\epsilon_i - \frac{1}{2}\sum_{p,q}^{N} \eta_{s_p}\eta_{s_q}\delta_{s_p s_i}\delta_{s_q s_i}\right] \tag{B3}$$

$$\times \delta\left( X_i(d) - g_{s_i}\eta_{s_i} - \sqrt{1-g_{s_i}^2}\epsilon_i \right). \tag{B4}$$

This is simplified to the following form, where the sum over $i$ stocks is converted to sums of the clusters $s$ and the $n_s$ stocks in each cluster

$$P = \prod_{s=1}^{S}\prod_{d=1}^{D} \int d\eta_s e^{-\frac{1}{2}\eta_s^2} \prod_{i\in s}^{n_s} \int d\epsilon_i \exp\left[ -\frac{1}{2}\epsilon_i^2 \right] \delta\left( X_i(d) - g_s\eta_s - \sqrt{1-g_s^2}\epsilon_i \right). \tag{B5}$$

The Gaussian integral over the delta function is evaluated relative to the $\epsilon_i$'s, using that $\prod \int f(x)\delta(ax - x_0) = \prod \frac{1}{|a|} f(x_0/a)$ over the $n_s$ delta functions,

$$P = \prod_{s=1}^{S}\prod_{d=1}^{D} \int \frac{d\eta_s}{(1-g_s^2)^{\frac{n_s}{2}}} e^{-\frac{1}{2}\eta_s^2} \prod_{i\in s}^{n_s} \exp\left[ -\frac{1}{2}\frac{(g_s\eta_s - X_i)^2}{1-g_s^2} \right]. \tag{B6}$$

Expanding out the integrand and using $\prod_i e_i^A = e^{\sum_i A_i}$,

$$P = \prod_{s=1}^{S} \prod_{d=1}^{D} \int \frac{d\eta_s}{(1-g_s^2)^{\frac{n_s}{2}}} \exp\left[ -\frac{1}{2}\eta_s^2 - \frac{1}{2}\sum_{i\in s}^{n_s} \frac{(g_s^2\eta_s^2 - 2g_s\eta_s X_i + X_i^2)}{1-g_s^2} \right]. \tag{B7}$$

Expanding out the sum terms and evaluating where possible

$$P = \prod_{s=1}^{S} \prod_{d=1}^{D} \int \frac{d\eta_s}{(1-g_s^2)^{\frac{n_s}{2}}} \exp\left[ -\frac{1}{2}\eta_s^2 - \frac{1}{2}\frac{n_s g_s^2 \eta_s^2}{1-g_s^2} + \frac{g_s\eta_s}{1-g_s^2}\sum_{i\in s}^{n_s} X_i - \frac{1}{2}\frac{1}{1-g_s^2}\sum_{i\in s}^{n_s} X_i^2 \right]. \tag{B8}$$

This can be further simplified to

$$P = \prod_{s=1}^{S} \prod_{d=1}^{D} \int \frac{d\eta_s}{(1-g_s^2)^{\frac{n_s}{2}}} \exp\left[ -\frac{1}{2}\frac{1-g_s^2+n_s g_s^2}{1-g_s^2}\eta_s^2 \frac{g_s\eta_s}{1-g_s^2}\sum_{i\in s}^{n_s} X_i - \frac{1}{2}\frac{1}{1-g_s^2}\sum_{i\in s}^{n_s} X_i^2 \right]. \tag{B9}$$

We now evaluate the Gaussian integral using that $\int e^{-x^2}dx = \sqrt{\pi/2}$ and hence that $\int e^{-ax^2+bx}dx = \sqrt{\frac{\pi}{2a}}e^{\frac{b^2}{4a}}$

$$P = \prod_{s=1}^{S} \prod_{d=1}^{D} \frac{\sqrt{\pi}}{(1-g_s^2)^{\frac{n_s}{2}}} \frac{(1-g_s^2)^{\frac{1}{2}}}{(n_s g_s^2 + (1-g_s^2))^{\frac{1}{2}}} \tag{B10}$$

$$\times \exp\left[ \frac{g_s^2}{2(n_s g_s^2 + (1-g_s^2))(1-g_s^2)} (\sum_{i\in s}^{n_s} X_i)^2 \right] \tag{B11}$$

$$\times \exp\left[ -\frac{1}{2}\frac{1}{1-g_s^2}\sum_{i\in s}^{n_s} X_i^2 \right]. \tag{B12}$$

Evaluating the product of all $D$ times, where $D >> 1$,

$$P = \prod_{s=1}^{S} \left[ \frac{\sqrt{\pi}}{(1-g_s^2)^{\frac{n_s}{2}}} \frac{(1-g_s^2)^{\frac{1}{2}}}{(n_s g_s^2 + (1-g_s^2))^{\frac{1}{2}}} \right]^D \tag{B13}$$

$$\times \exp\left[ \frac{g_s^2}{2(n_s g_s^2 + (1-g_s^2))(1-g_s^2)} (\sum_{d}^{D}\sum_{i\in s}^{n_s} X_i)^2 \right] \tag{B14}$$

$$\times \exp\left[ -\frac{1}{2}\frac{1}{1-g_s^2}\sum_{d}^{D}\sum_{i\in s}^{n_s} X_i^2 \right]. \tag{B15}$$

Using that $C_{ij} = \frac{1}{D}\sum_d X_i X_j$,

$$\sum_{d}^{D} (\sum_{i\in s} X_i)^2 = \sum_{i,j=1}^{N} (\sum_{d}^{D} X_i X_j)\delta_{s_i s}\delta_{s_j,s} = Dc_s, \tag{B16}$$

and that the variance of the process in the $s^{th}$ cluster can be computed from the trace[1]

$$\sum_{i\in s}\sum_{d}^{D} X_i^2 = DC_{ii} = \sum_{i\in s}^{n_s} DC_{ii} = Dn_s. \tag{B18}$$

Substituting Equation B16 and B18 into Equation B15,

$$P = \prod_{s=1}^{S}\left[\frac{\sqrt{\pi}}{(1-g_s^2)^{\frac{n_s}{2}}}\frac{(1-g_s^2)^{\frac{1}{2}}}{(n_s g_s^2+(1-g_s^2))^{\frac{1}{2}}}\right]^D \exp\left[-\frac{D}{2}\frac{n_s}{1-g_s^2}+\frac{D}{2}\frac{c_s}{1-g_s^2}\frac{g_s^2}{n_s g_s^2+(1-g_s^2)}\right] \tag{B19}$$

We can rewrite this as

$$P = \prod_{s=1}^{S}\frac{\pi^{\frac{D}{2}}(n_s g_s^2+(1-g_s^2))^{\frac{-D}{2}}}{(1-g_s^2)^{\frac{D}{2}(n_s-1)}}\exp\left[-\frac{D}{2}\frac{1}{1-g_s^2}\left(n_s-\frac{c_s g_s^2}{n_s g_s^2+(1-g_s^2)}\right)\right]. \tag{B20}$$

Then using that $P \propto e^{-DH_c}$, we can find $H_c \propto \ln(P)$ from Equation B20, and using that $\ln\prod_i A_i = \sum_i \ln(A_i)$ to find the log-likelihood function [Need to use $D >> 1$ and look at expansion $(g_s - g_s^*)$]

$$\ln(P) = -\frac{D}{2}\sum_{s=1}^{S}\left[\ln(n_s g_s^2+(1-g_s^2))\right. \tag{B21}$$

$$+ \quad \left.(n_s-1)\ln(1-g_s^2)\right] \tag{B22}$$

$$+ \quad \frac{D}{2}\sum_{s=1}^{S}[\ln(\pi)] \tag{B23}$$

$$- \quad \frac{D}{2}\sum_{s=1}^{S}\frac{1}{1-g_s^2}\left[n_s-\frac{c_s g_s^2}{n_s g_s^2+(1-g_s^2)}\right]. \tag{B24}$$

Using Equation A11, we can substitute for $g_s$ in A11 to find the log-likelihood entirely in terms of $n_s$ and $c_s$, using that $(1-g_s^2) = \frac{n_s^2-c_s}{n_s^2-n_s}$ and $\frac{c_s}{n_s} = n_s g_s^2 + (1-g_s^2)$:

$$H_c = \frac{1}{2}\sum_{s:n_s>0}\left[\log\frac{c_s}{n_s}+(n_s-1)\log\frac{n_s^2-c_s}{n_s^2-n_s}\right]+\frac{1}{2}\sum_{s:n_s>0}\left[\ln(\pi)+n_s\right]. \tag{B25}$$

The last term is a constant, given that $\sum_{s:n_s>0} n_s = N$ where $N$ is the number of objects. This is fixed for a given system. Hence the likelihood function required is

$$H_c = \frac{1}{2}\sum_{s:n_s>0}\left[\log\frac{c_s}{n_s}+(n_s-1)\log\frac{n_s^2-c_s}{n_s^2-n_s}\right] \tag{B26}$$

up to a constant $\frac{1}{2}(S\ln(\pi)+N)$.

[1] The trace of the correlation matrix for each cluster s can be verified from the eigenvalues

$$\sum_{i}^{N} C_{ii} = \sum_{s}\lambda_s = (n_s-1)(1-g_s^2)+n_s g_s^2+(1-g_s^2) = n_s. \tag{B17}$$